\documentclass[10pt,times,3p]{elsarticle}

\usepackage{graphicx}
\usepackage[T1]{fontenc}    
\usepackage{url}            
\usepackage{booktabs}       
\usepackage{amsfonts}       
\usepackage{amsmath}
\usepackage{amsthm}
\usepackage{nicefrac}       
\usepackage{microtype}      
\usepackage{xcolor}         

\usepackage[textfont=it,labelfont=bf]{caption} 
\usepackage{subcaption}
\usepackage{algorithm}
\usepackage{algpseudocode}
\usepackage{shadethm}
\usepackage{thmtools}
\usepackage{pifont}

\usepackage{colortbl}
\usepackage{multirow}
\usepackage[para,online,flushleft]{threeparttable}
\usepackage{array}
\usepackage{tabularray}
\usepackage{enumitem}

\usepackage{pgfplots}
\pgfplotsset{compat=1.17}

\usepackage{hyperref}       
\hypersetup{
    breaklinks=true,
    colorlinks,
    linkcolor={blue!80!black},
    citecolor={blue!80!black},
    urlcolor={blue!80!black},
    plainpages=true
}
\newcommand{\cref}[2]{\hyperref[#2]{#1~\ref*{#2}}}
\newcommand{\colref}[3]{\hyperref[#2]{#1~\ref*{#2}{#3}}}
\newcommand{\figref}[1]{\cref{Figure}{#1}}

\newcommand{\secref}[1]{\cref{Section}{#1}}
\newcommand{\eqnref}[1]{\cref{Equation}{#1}}
\newcommand{\tabref}[1]{\cref{Table}{#1}}
\newcommand{\algoref}[1]{\cref{Algorithm}{#1}}

\declaretheoremstyle[%
  spaceabove=-6pt,%
  spacebelow=6pt,%
  headfont=\bfseries\itshape,%
  postheadspace=0.5em,%
  qed=\qedsymbol%
]{mystyle}

\theoremstyle{mystyle}

\newcommand{\vbv}{{VbV}}
\newcommand{\Active}{\textsc{Active}}
\newcommand{\InActive}{\textsc{InActive}}
\newcommand{\petsc}{\href{https://www.mcs.anl.gov/petsc/}{PETSc}}

\definecolor{trueintercepted}{HTML}{FFFF00}
\definecolor{falseintercepted}{HTML}{F08080}
\definecolor{trueboundary}{HTML}{00007f}
\definecolor{notintercepted}{HTML}{008000}

\journal{Finite Elements in Analysis \& Design}
\begin{document}

\begin{frontmatter}
\title{High-Resolution Thermal Simulation Framework for Extrusion-based Additive Manufacturing of Complex Geometries}

\author[ISU]{Dhruv Gamdha}
\author[ISU]{Kumar Saurabh}
\author[ISU]{Baskar Ganapathysubramanian\texorpdfstring{\corref{cor1}}{}}
\author[ISU]{Adarsh Krishnamurthy\texorpdfstring{\corref{cor1}}{}}
\affiliation[ISU]{organization={Iowa State University}, 
            city={Ames},
            state={Iowa},
            country={USA}}
\cortext[cor1]{Corresponding Authors}

\begin{abstract}

Accurate simulation of the printing process is essential for improving print quality, reducing waste, and optimizing the printing parameters of extrusion-based additive manufacturing. Traditional additive manufacturing simulations are very compute-intensive and are not scalable to simulate even moderately sized geometries. In this paper, we propose a general framework for creating a digital twin of the dynamic printing process by performing physics simulations with the intermediate print geometries. Our framework takes a general extrusion-based additive manufacturing G-code, generates an analysis-suitable voxelized geometry representation from the print schedule, and performs physics-based (transient thermal) simulations of the printing process. Our approach leverages adaptive octree meshes for both geometry representation as well as for fast simulations to address real-time predictions. We demonstrate the effectiveness of our method by simulating the printing of complex geometries at high voxel resolutions with both sparse and dense infills. Our results show that this approach scales to high voxel resolutions and can predict the transient heat distribution as the print progresses. Because the simulation runs faster than real print time, the same engine could, in principle, feed thermal predictions back to the machine controller (e.g., to adjust fan speed or extrusion rate). The present study establishes the computational foundations for a real-time \emph{digital twin}, which can be used for closed control loop control in the future.

\end{abstract}

\end{frontmatter}

\section{Introduction}
\label{Sec:Introduction}

\begin{figure*}[!ht]
    \centering
    \includegraphics[trim={0 6.5cm 0 4.5cm},clip,width=0.99\linewidth]{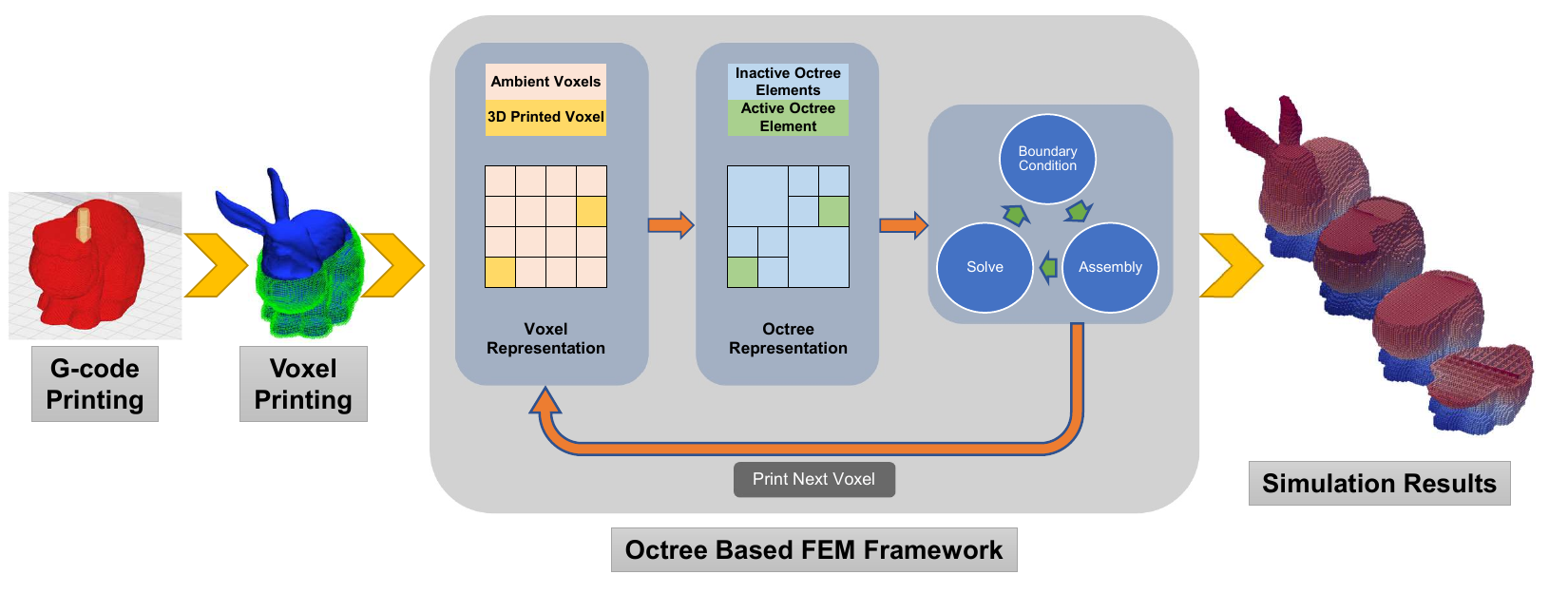}
    \caption{ Illustration of the proposed voxel-based, octree-enabled FEM framework for real-time 3D-printing simulations. Starting from a standard G-code model (left), the part is converted into a voxel representation where each newly deposited voxel (green) expands the active region. An octree mesh (center) then refines or coarsens elements based on print progression, enabling efficient assembly and solution of the evolving thermal problem. Finally, the simulation iterates this deposition-and-solve process until the full 3D object (right) is printed. }
    \label{Fig:Flowchart}
\end{figure*}

 Fused deposition modeling (FDM) is an additive manufacturing process that builds 3D objects layer-by-layer by melting and extruding thermoplastic filaments through a heated nozzle. The nozzle moves along a predetermined path, depositing the molten material in precise locations to create the desired shape. Once deposited, the material then solidifies to create the final part. FDM is a popular and widely used technology due to its low cost, versatility, and ease of use. It can produce parts with good mechanical properties and accuracy. Such 3D printing approaches have revolutionized the manufacturing industry by enabling the production of complex geometries with high precision and customization. However, FDM faces a few challenges in achieving the desired print quality and reducing material waste, especially for complex geometries with sparse infills. One challenge is the need for support structures for overhanging features or complex geometries, which can increase material waste and post-processing time. Another is the possibility of warping and distortion due to thermal stresses during printing, which can affect the final dimensional accuracy of the part.

Accurate simulation of the 3D printing process can help address these challenges by predicting the thermal and mechanical stress distribution as the build progresses, which can inform print quality and material usage. Fast and accurate simulations can help rapidly explore the print process and design non-trivial approaches that minimize material usage while ensuring that thermal stresses are kept to acceptable levels. The ability of real-time simulations also opens up the possibility of creating digital twins of the 3D print process. These digital twins can assimilate measurements from the physical build and accurately predict the intermediate and final shapes. Additionally, once the computed temperature field can be returned in near-real-time, the same framework could drive model-predictive adjustments to fan speed, extrusion rate, or chamber temperature, thereby closing the control loop.

Simulating the 3D printing process is challenging for several reasons. These include computational challenges associated with (a) efficiently representing complex print geometries, (b) the multiscale nature of the process (small time scales of the nozzle-print interaction vs. long simulation time horizon of the entire print), (c) the coupled multi-physics phenomena (thermal, phase-change, mechanical), and (d) the time- and location-dependent material properties of the print, as well as the complexities of the print schedule (intricate infill patterns, variable resolution, variable material feeds). A final challenge is to simulate the time-dependent printing process in print-time, i.e., faster than the physical print process, to enable control and fast design exploration. 

Simulating 3D printing remains a very active research area, with several approaches that resolve some (but not all) of these challenges. Recently, voxel-based methods have emerged as a promising approach for simulating 3D printing, with the possibility of resolving all the challenges listed above. In voxel-based methods, the geometry is represented as a collection of voxels (i.e., volumetric pixels) that capture the printing order (and infill patterns) and material properties. Voxel-based simulations have the advantage of being computationally efficient and flexible in representing complex geometries with arbitrary infill patterns, and the structured representation offers the possibility of fast simulations. In this paper, we propose a novel framework for voxel-by-voxel 3D printing simulation that leverages adaptive octree meshes to efficiently represent the geometry and ensure highly optimized physics simulations. Our framework consists of three distinct stages, as shown in \figref{Fig:Flowchart}:
\begin{itemize}
    \item Converting GCode into an intermediate voxel representation. We first convert the GCode into a voxel-by-voxel (abbreviated as \textit{VbV}) print schedule. This intermediate representation is flexible enough to account for sparse and dense infill patterns.
    \item Converting the intermediate voxel representation into an analysis-suitable octree representation, with graceful adaptive coarsening and refinement as the print schedule progresses.
    \item A scalable FEM simulator based on adaptive octree meshes that predicts transient thermal fields during the print process.
\end{itemize}
Although the present paper demonstrates only the forward (simulation) path, the achieved speed-ups mean that the solver output is already fast enough to be consumed by an external controller in a future closed-loop implementation. We first compare our framework with experimental data for single filament rectangular geometry and, finally, demonstrate the effectiveness of our framework using several complex geometries at high voxel resolution and under different infill conditions. To the best of our knowledge, our approach is the first to produce high-fidelity, real-print-time simulations of the 3D printing process for complex geometries by leveraging parallel adaptive octree meshes for FEM-based time-dependent 3D simulations.

The remainder of this paper is structured as follows: in \secref{Sec:RelatedWork}, we review the existing literature on 3D printing simulations, voxel-based representations, and physics-based modeling. \secref{Sec:Methodology} describes the methodology of our approach, including the details of each of the three stages. \secref{Sec:Results} presents the results of our simulations and compares them to existing methods or benchmarks, if available. Finally, we discuss the strengths and weaknesses of our approach and the implications of our results for the field of 3D printing simulations in \secref{Sec:DiscussionConclusion}.

\section{Background and Related Work}
\label{Sec:RelatedWork}

Fused Deposition Modeling (FDM) is a widely used additive manufacturing technology that has gained immense popularity in the last decade due to its low cost, ease of use, and ability to produce complex geometries \citep{chennakesava2014fused, grimm2003fused}. The process involves extruding a thermoplastic material layer by layer until the final object is formed. FDM technology has found its applications in various fields, including the production of metal \citep{mireles2012fused}, ceramics \citep{bellini2005new}, polymers \citep{carneiro2015fused, penumakala2020critical}, and composites \citep{rahim2019recent, ning2015additive}. One of the notable advantages of FDM technology is the ability to create parts with varying densities, which can have significant implications for various applications \citep{dev2021effect}. Another area where FDM has found its use is in the medical industry, where it is used for manufacturing patient-specific implants and drugs \citep{espalin2010fused, dumpa20213d}. FDM technology has also gained popularity in the aerospace industry \citep{ning2015additive}.

Despite its advantages, FDM printing faces several challenges that can affect the quality and mechanical properties of the printed parts. One of the most common issues is the surface quality, which can be affected by the stair-step effect due to the layer-by-layer deposition process \citep{pandey2003improvement, parulski2021challenges}. Another challenge is achieving the desired surface roughness, which can be influenced by factors such as the nozzle diameter, layer height, and print speed \citep{vyavahare2020experimental}. Furthermore, the mechanical properties of FDM printed parts can be compromised due to weak interlayer bonding and the high melt viscosity of some materials, such as thermoplastic elastomers \citep{awasthi2021fused}. Achieving high print speeds can also be challenging, resulting in reduced mechanical strength and lower part accuracy \citep{wang2019effects}. These challenges have motivated research efforts to improve FDM printing technology and optimize the process parameters for better part quality and performance.

Experimental analysis of FDM parts is essential to understand their quality, behavior, and performance in different applications. Mechanical testing, dimensional analysis, surface roughness analysis, internal defect characterization, and microstructural analysis are commonly used experimental methods \citep{krolczyk2014experimental, galantucci2015analysis, venkatraman2021experimental, gamdha2021automated}. However, experimentally analyzing FDM parts poses several challenges. One of the major challenges is the interplay of various process parameters, such as temperature, layer thickness, infill percentage, and print speed. These parameters affect the quality and performance of the printed parts, making it difficult to obtain consistent and accurate results \citep{sheoran2020fused}. The inherent anisotropy of FDM parts is another challenge, as their properties vary in different directions, making it challenging to obtain representative test results \citep{ahn2002anisotropic}. Additionally, FDM parts may contain voids, inclusions, or other defects due to the nature of the process, which can impact their performance and hinder the accuracy of test results \citep{10.1115/1.2830582}. The size and shape of the sample can also affect the test results, making it challenging to compare different samples or generalize findings to other parts or applications.

Computational tools can play a crucial role in addressing the challenges faced by experimentalists in analyzing FDM parts. For example, simulation software can model the FDM process and optimize the process parameters for specific applications \citep{ozen2021optimization}. This can help to reduce the number of experimental trials needed and provide insights into the effects of different process parameters on the quality and performance of the printed parts. In addition, computational methods such as finite element analysis (FEA) and computational fluid dynamics (CFD) can be used to predict the mechanical behavior of FDM parts, including their strength, stiffness, and deformation.

There have been a few physics-based computational studies of the additive manufacturing process. \citet{ben2022residual} showed that residual stresses in additive manufacturing parts result from thermal expansion during heating, contraction during cooling, and volume change due to the phase transformations. The presence of these stresses can be catastrophic. \citet{bayat2021review} shows the difficulty in performing full-body FEM simulations of the additive manufacturing process, which incorporates relevant physical phenomenons such as fluid flow, temperature, and stress evolution due to very high computational requirements and the inability of the models to scale up\citep{khara2024neural, shah2022gpu, shadkhah2025octree, rabeh2024modeling, khara2024solving}. \citet{chen2020high} and \citet{bailey2017laser} have shown a coupled simulation of CFD and solid mechanical model but could only simulate two tracks due to high computational requirements. The lack of computational tools makes the analysis of the effect of these parameters on the final part very difficult. Our work helps overcome many computational challenges because of the easy conversion of GCode into the voxel representation, efficient mesh creation approach, and scalable FEM framework.

\citet{vovrivsek2023gpams} presents a GCode processor for advanced additive manufacturing simulations. The paper focuses on the creation of voxel geometry from GCode, which can be used for Finite Element Method (FEM) simulations. However, their simulation is limited to the voxel geometry creation, and they do not perform any FEM simulation, which, in our understanding, is a lot more complex and computationally expensive. Additionally, the resolution of their high-resolution geometry is comparable to ours. \citet{baiges2021adaptive} presents an adaptive Finite Element strategy for the numerical simulation of additive manufacturing processes. The paper uses an adaptive octree mesh that is fine only in regions of high change and coarse elsewhere. The paper shows FEM-based steady-state mechanical analysis performed only on simple geometries like cuboids. The number of time steps they simulate is limited to only 100, and the total time taken is around 500 minutes. In comparison, our work performs FEM simulations over the entire printing process, using a more efficient octree representation. Our FEM framework can handle more complex geometries, as we have demonstrated with examples like the Stanford Bunny and Moai. Being a fast, accurate, and scalable FEM framework, we demonstrate the potential of our framework by simulating the complete printing of the Stanford Bunny and Moai, as well as the future potential of performing coupled thermo-mechanical simulations. Our total time steps solved for is much higher than the other two papers, with total mesh nodes ranging up to 350K to 500K. Additionally, our print geometry update resolution is at the voxel level, and we are able to simulate the full body of complicated geometries, not just the final mesh.

Several other authors have tackled the FDM heat-transfer problem with simplified physics but different numerical settings. \citet{ramos2022experimental} calibrated convection coefficients for ABS and then used an \textsc{ANSYS} element-activation strategy combined with element homogenisation to accelerate cooling-stage analysis. \citet{cattenone2019finite} developed a thermo-mechanical FE framework in \textsc{Abaqus} and demonstrated residual-stress prediction for simple ABS beams, while \citet{xu2021thermal} coupled in-situ thermocouple data with a porosity-corrected thermal model to study PLA thin walls. All three studies start from pre-meshed CAD geometry and therefore still require manual mesh preparation and path scripting, in contrast to the G-code-driven, voxel-to-octree pipeline proposed here.

A closely related effort is the heat-transfer study of \citet{ramos2023efficient}, who accelerate FFF simulations inside a commercial \textsc{ANSYS} environment by periodically remeshing and coarsening regions that have cooled.  Their workflow, however, starts from a
pre-meshed solid and a hand-scripted element-activation sequence; the paper does not discuss how the nozzle path is extracted from slicer output, nor does it maintain separate air and polymer elements once a region is homogenised. In contrast, the present work ingests raw G-code, converts it to a voxel representation, builds a layer-synchronous adaptive octree \emph{without user parameters}, and keeps air and polymer voxels disjoint throughout the build. These choices allow spatially varying convection coefficients, preserve interface fidelity, and enable real-time simulation of highly complex, sparse-infill geometries directly from the printer’s input file.

\section{Additive Manufacturing Simulation Framework}
\label{Sec:Methodology}

In this section, we present a voxel-based 3D printing simulation framework that addresses the challenges associated with fast simulations of the FDM process. This simulation framework consists of a hierarchical and adaptive representation of the printing geometry and a physics-based simulation of the printing process. Our method leverages the power of hierarchical octree data structures and adaptive mesh refinement and coarsening strategy to efficiently generate a dynamic analysis-suitable geometry representation that changes with each new voxel print. The framework is designed to be scalable and can be used to simulate the printing process for complex geometries at high voxel resolutions. The framework can also be used to simulate the printing process in real-time, i.e., faster than the physical printing process, to enable control and fast design exploration. \figref{Fig:Flowchart} shows the flow chart of the proposed simulation method. The framework consists of three main stages: (1) converting the GCode into an intermediate voxel representation, (2) converting the voxel representation into an analysis-suitable geometry representation, and (3) performing physics-based simulations of the printing process. The following sections describe each of these stages in detail.

\subsection{Voxel-by-Voxel (VbV) Printing Order}
\label{SubSec:VoxelPrintingOrder}

\begin{figure*}[!t]
    \centering
    \includegraphics[width=0.99\linewidth, trim={2.0cm 0.8cm 2.0cm 0.8cm}, clip]{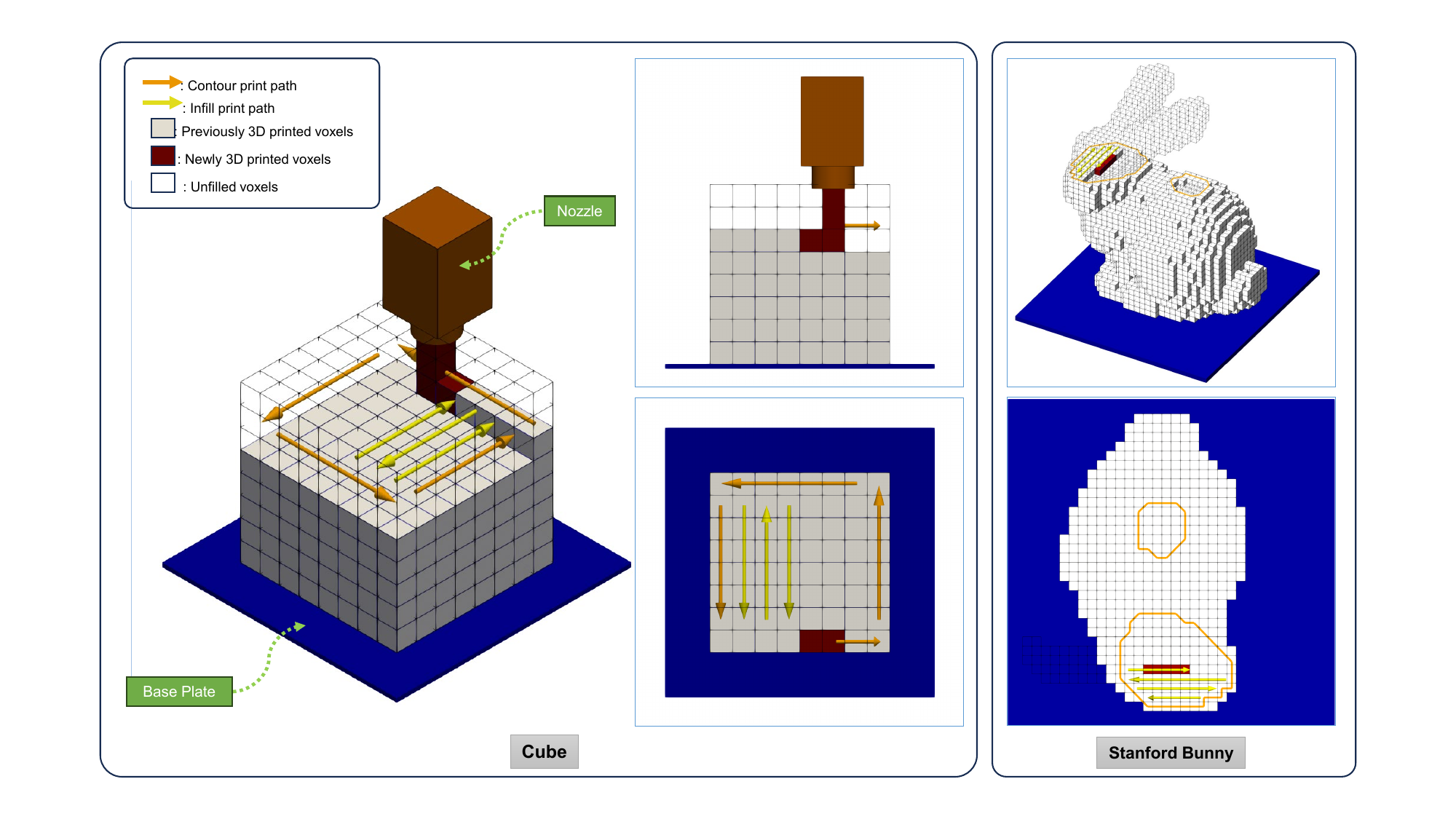}
    \caption{Voxel-by-voxel printing order for the cube and Stanford bunny geometry. The figure shows the 3D printing of the voxel geometry. The red voxels are the ones that have been printed recently, gray voxels are the ones that have been printed previously, and the white voxels are the ones that are yet to be printed. Orange arrow shows the contour print direction and the yellow arrow shows the infill print direction.}
    \label{Fig:VbV_printing_cube_bunny}
\end{figure*}

The first step in our simulation method involves creating a voxel-by-voxel (\vbv{}) printing order from the GCode. A \vbv{} printing order is a list of voxels in the order in which they are printed. The \vbv{} printing order is a voxelized representation of the GCode and is created by projecting the toolpath onto a uniform Cartesian grid, which we call the voxel grid ($V_G$). We start with the GCode file and project the toolpath onto a uniform Cartesian grid. The toolpath information is extracted from the GCode and is converted into a series of line segments corresponding to the extruder's movement during printing. The voxels that are intersected by these line segments are then marked as active. This provides us with a uni-level voxel grid representation of the geometry. The grid resolution is dictated by voxel size, representing the smallest unit of the grid. We then use the method proposed by \citet{GHADAI2021101929} to generate the layer-by-layer contours and infill patterns. These patterns are then converted into a \vbv{} printing order. The \vbv{} order serves as input for our subsequent geometry generation and physics-based simulation modules, enabling us to predict the temperature distribution during the printing process. \figref{Fig:VbV_printing_cube_bunny} shows the 3D printing of the voxel geometry according to the \vbv{} printing order for the cube and Stanford bunny geometry.

\subsection{Analysis Suitable Adaptive Octree Mesh}
\label{SubSec:AnalysisSuitableAdaptiveOctreeMesh}
Adaptive octree mesh has been widely used in computational sciences because of its simplicity and ability to scale to a large number of processors~\citep{rudi2015extreme,sundar2008bottom,saurabh2021scalable,ishii2019solving,burstedde2011p4est}. Specifically, we used \textit{2:1 balanced axis aligned linearized} octree. An octree is said to be \textit{2:1 balanced} if two neighboring octants do not differ by more than 1 level (i.e., neighboring octants can only differ in size by a factor of $1/2$). An \textit{axis-aligned} octree means that each element of the octree has its axis parallel to the Cartesian coordinate axis; in other words, the elements are not deformed. A linearized octree means that each element of the mesh can be represented by the leaf of an octree, which can be linearized into an array. The linear array is obtained by traversing the octree traversal in Morton or Hilbert order. Such a traversal allows for a good locality for the computations\citep{bader2012space}. These features make octree suitable for performing the analysis relevant to the current work.

\subsubsection{Voxel and Octree Grids}
\label{SubSec:Grids}

\begin{figure*}[!ht]
    \centering
    \includegraphics[width=0.99\linewidth, trim={0cm 3.0cm 0cm 3.0cm}, clip]{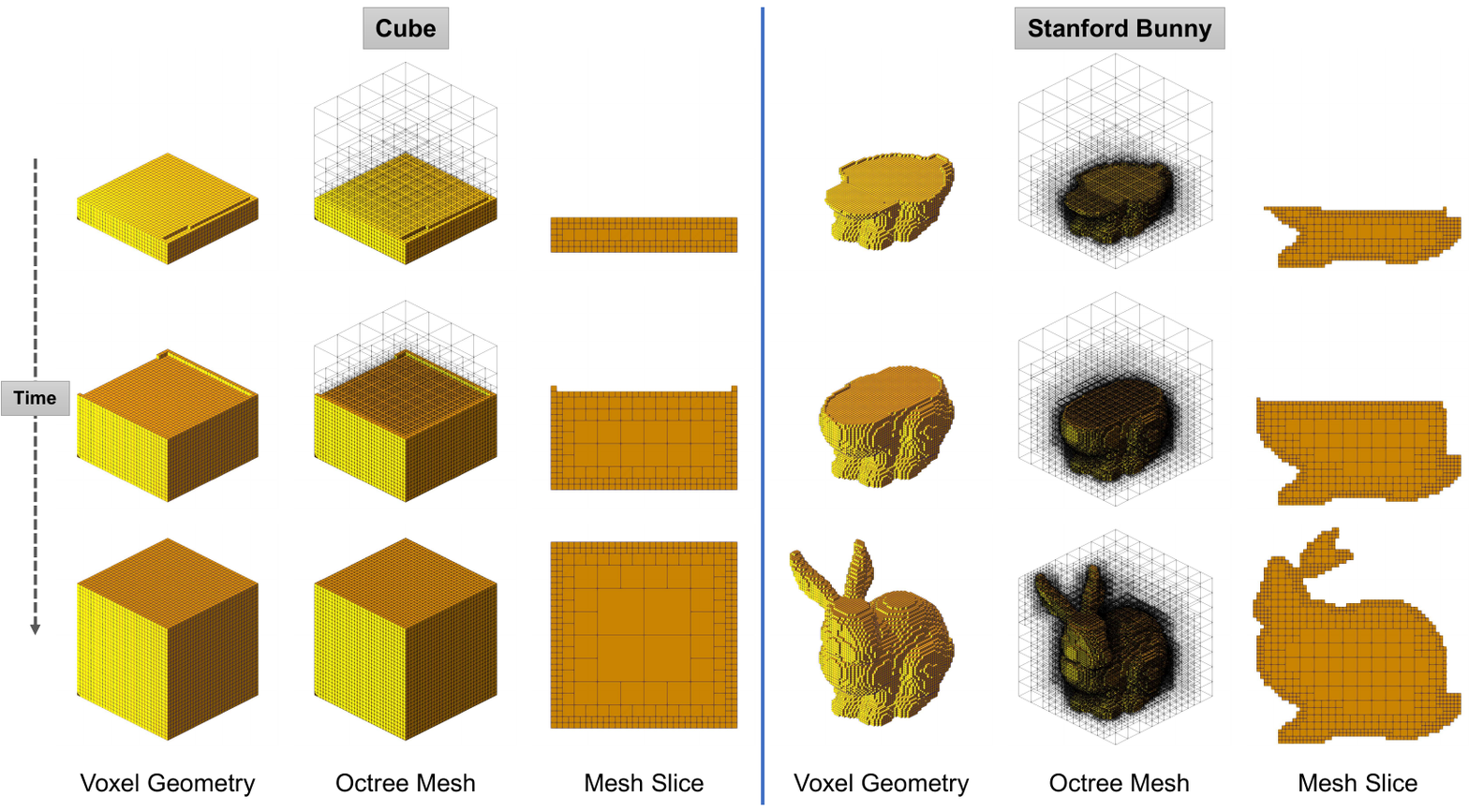}
    \caption{ Voxel-based geometry, octree mesh, and 2D slices at three stages of the build for both a cube (left) and the Stanford Bunny (right). From top to bottom, each row represents an increasingly complete print. The left column in each block shows the voxel geometry at the given time, the center column shows the corresponding octree mesh (with inactive elements in wireframe), and the right column presents a mesh slice for clarity. This progression illustrates how the geometric representation, meshing strategy, and cross-sectional patterns evolve as additional layers are deposited.}
    \label{Fig:voxelMeshSlice}
\end{figure*}

In our simulation method, we utilize two distinct grid types: the voxel grid ($V_G$) and the octree grid ($O_G$) as shown in \figref{Fig:voxelMeshSlice}. The voxel grid, $V_G$, is a uniform Cartesian grid central to the \vbv{} printing process. We adopt a non-dimensional framework where each voxel's dimension, $\Delta x_i$, corresponds to one non-dimensional unit. A voxel $v \in V_G$ is deemed 'active' once printed and participates in all subsequent simulation steps.

The finite element method (FEM) simulations, however, are conducted on the octree grid, $O_G$. This grid complements the voxel grid's \vbv{} process but introduces variable-sized elements (or octants), allowing for computational adaptability. Specifically, octants near newly active voxels are finely refined, while those in less critical areas are coarser, thereby optimizing computational resources. In all simulations, the octant corresponding to a newly activated voxel is refined to ensure that $\Delta x_{O_G} \leq \Delta x_{V_G}$. 

To effectively track the evolving structure during printing, we employ a bitset vector $V_B$ within the voxel grid $V_G$. This vector, sized at $N_{V_G}$ (the total voxel count in $V_G$), uses 7 bits per voxel. The first bit signifies the voxel's active status, while the remaining six represent the state of each of its six faces, indicating boundary presence. As printing progresses, this bitset is dynamically updated. Activating a voxel sets its first bit to 1 and triggers a check on its six neighboring voxels. If a neighbor is inactive, the corresponding voxel face is marked as a boundary (bit set to 1); if active, the shared face is no longer a boundary (bit set to 0). These updates are critical for constructing the octree grid $O_G$, ensuring a detailed and responsive representation of the printed structure.

\subsubsection{Octree Construction Using Refinement and Coarsening}
\label{SubSec:OctreeConstruction}

The construction of our octree grid is a pivotal process in our simulation, essential for dynamically adjusting the grid resolution according to the needs of the simulation. This construction starts at the highest level with a root cube and progresses in a top-down manner. Specifically, when a voxel $v$ is activated at a new height $h$ in the voxel grid $V_G$, the octree grid elements on the x-y plane at this height undergo refinement. By refining the entire x-y plane at the new height, we ensure that all elements at this level are uniformly refined to accommodate the new voxel. Using this apprach, we perform refinement of $O_G$ only once for each new height level in $V_G$. This strategy is particularly efficient as it avoids the need to refine the entire octree grid at every voxel activation, which would be computationally expensive and unnecessary.

The refinement decision for each element is determined by its spatial relationship with the newly activated voxel. We calculate the equivalent integer coordinates in $V_G$ for the nearest and farthest nodes of an octant, checking whether the height range of the octant intersects with that of the activated voxel. Elements are then flagged as either \textsc{Refine} or \textsc{No Change} based on this assessment. Elements flagged for \textsc{Refine} undergo an increase in resolution, dividing into eight smaller octants, whereas those flagged as \textsc{No Change} maintain their current level. This process of refinement is recursively repeated, following the flagging procedure, until all elements achieve their required levels, indicated when all flags are \textsc{No Change}. The algorithmic steps for this octree construction process, which balances computational efficiency with the need for precision in the simulation, are detailed in \algoref{alg:octreeRefine}.

\begin{algorithm}[t!]
\footnotesize
    \caption{\textsc{OctreeRefine:} \footnotesize{Octree Refinement procedure}}
    \label{alg:octreeRefine}
    \begin{algorithmic}[1]
\Require $V_G$, $v$ (activated voxel), $O_G$, $O_L$(Level of octree for required voxel)
\Ensure Refined octree
\item[]
\State id\_3d :\texttt{int} $\leftarrow$ compute\_3d\_id($v, V_G$) \Comment{Compute the voxel id in 3D coordinate for voxel $v$}.
\State $R_{flags} \leftarrow$ [\textsc{No Change}]  \Comment{Vector for refine flags filled with \textsc{No Change}}
\For{ $e \in O_G$ }
\State id\_min :\texttt{int}$\leftarrow$ compute\_3d\_id($e_{min}, V_G$) \Comment{Compute the voxel id in 3D coordinate for voxel $e_{min}$}.
\State id\_max :\texttt{int}$\leftarrow$ compute\_3d\_id($e_{max}, V_G$) \Comment{Compute the voxel id in 3D coordinate for voxel $e_{max}$}.
\If{id\_min[$2$] $\leq$ id\_3d[$2$] $\leq$ id\_max[$2$]}  \Comment{Checking if octree element height intersects with the voxel}
\If {level {$(e) \leq O_L$}} \Comment{Checking the octree RefineFlags}
\State $R_{flags}[e] \leftarrow $ \textsc{Refine} 
\EndIf
\EndIf
\EndFor
\If{all\_of[R] == \textsc{No Change}} \Comment{No change required for octree}
\State \Return $O_G$
\Else 
\State $O_G \leftarrow $ construct($R_{flags}$) \Comment{Construct octree with refine flags and transfer solution}
\State \Return \textsc{OctreeRefine($V_G,v,O_G,O_L)$}\Comment{Recursively refine till all the elements are at required level}
\EndIf
    \end{algorithmic}
\end{algorithm}

In our octree grid, coarsening is an essential process to optimize computational efficiency. This process involves the recursive merging of child octants within a parent octant, provided they are at the same level and share identical material properties (i.e., either all active or inactive). For an octant to be eligible for merging, it must satisfy the following conditions:
\begin{enumerate}
    \item The octant is not a boundary octant, ensuring that any boundary conditions are not violated.
    \item Its refinement level is above the base level, indicating it has undergone prior refinement.
    \item It does not reside at the current printing height, ensuring the highest refinement in the active printing layer.
\end{enumerate}
An octant that meets these criteria is flagged for coarsening (\textsc{Coarsen}), while others are marked as \textsc{No Change}. This coarsening process continues recursively until all octants are flagged as \textsc{No Change}, ensuring that the grid is optimally simplified without compromising the simulation's accuracy or resolution. The detailed algorithmic steps of this coarsening process are outlined in \algoref{alg:octreeCoarsen}.

\begin{algorithm}[t!]
\footnotesize
    \caption{\textsc{OctreeCoarsen:} \footnotesize{Octree Coarsening procedure}}
    \label{alg:octreeCoarsen}
    \begin{algorithmic}[1]
\Require $V_G$, $V_B$, $v$ (activated voxel), $O_G$, $O_{BLvl}$(Base Level of octree)
\Ensure Coarsened octree
\item[]
\State id\_3d :\texttt{int} $\leftarrow$ compute\_3d\_id($v, V_G$) \Comment{Compute the voxel id in 3D coordinate for voxel $v$}.
\State $R_{flags} \leftarrow$ [\textsc{No Change}]  \Comment{Vector for coarsen flags filled with \textsc{No Change}}
\For{ $e \in O_G$ }
\State id\_min :\texttt{int}$\leftarrow$ compute\_3d\_id($e_{min}, V_G$) \Comment{Compute the voxel id in 3D coordinate for voxel $e_{min}$}.
\State id\_max :\texttt{int}$\leftarrow$ compute\_3d\_id($e_{max}, V_G$) \Comment{Compute the voxel id in 3D coordinate for voxel $e_{max}$}.

\State not\_bdry :\texttt{bool}$\leftarrow$ is\_not\_bdry($V_B[id\_min]$) \Comment{Check if the voxel is not a boundary voxel}

\State refined :\texttt{bool}$\leftarrow$ level($e$) $>$ $O_{BLvl}$ \Comment{Check if the octree element is already refined}

\State not\_current :\texttt{bool}$\leftarrow$ is\_not\_currentHeight($e$) \Comment{Check if the octree element is not located at current printing height}


\If{not\_bdry \textbf{and} refined \textbf{and} not\_current}  \Comment{Checking if octree element meets the coarsening criteria}
\State $R_{flags}[e] \leftarrow $ \textsc{Coarsen}
\EndIf

\EndFor
\If{all\_of[R] == \textsc{No Change}} \Comment{No change required for octree}
\State \Return $O_G$
\Else 
\State $O_G \leftarrow $ construct($R_{flags}$) \Comment{Construct octree with coarsen flags and transfer solution}
\State \Return \textsc{OctreeCoarsen($V_G,V_B,v,O_G,O_{BLvl})$}\Comment{Recursively coarsen till all the elements are at required level}
\EndIf
    \end{algorithmic}
\end{algorithm}

In the current implementation, we allow the octree to be refined or coarsened by a single level per iteration. One could alternately use multi-level refinement~\citep{saurabh2022scalable}, where each octant is refined directly to a required level. We are currently exploring the computational trade-off (solve time vs. refinement cost) of this multi-level refinement strategy.

\pagebreak

\subsubsection{Element Classification}
\label{SubSec:ElementClassification}

After constructing the octree grid $O_G$, we categorize each element in the mesh into two main types: \Active{} and \InActive{}. Elements classified as \Active{} overlap with the printed voxel region at a given time $t$, representing the physical part of the object being printed. Conversely, \InActive{} elements represent the unprinted volume, essentially the surrounding air. Further, we sub-categorize \Active{} elements into \textsc{In} and \textsc{Boundary} elements. \textsc{Boundary} elements have at least one face on the object's outermost layer, where we apply the Robin boundary condition to model heat transfer between the object and the air. On the other hand, \textsc{In} elements are internal and not on this boundary.

We denote the domain of \Active{} elements as $\Omega_A$. Therefore, $\Omega_I$, representing the \InActive{} domain, is defined as $\Omega_O - \Omega_A$, where $\Omega_O$ is the entire cubic space defined by the root-level octree. The classification into \Active{} and \InActive{}, as well as into \textsc{In} and \textsc{Boundary}, dynamically changes over time with the addition of new voxels $v$ to the system, altering $\Omega_A$ and $\Omega_I$ correspondingly.

It is important to note that only \Active{} regions contribute to the solution during the simulation. \InActive{} regions are excluded, typically by applying a Dirichlet boundary condition, effectively removing these areas from consideration during the solve step. This approach is similar to immersed-boundary methods as discussed in~\citep{saurabh2021industrial}. While dynamic matrix contraction/expansion could be used for adding or removing degrees of freedom, this is not preferred due to the significant memory overhead associated with dynamically managing contiguous memory.

\subsection{Thermal Simulations}
\label{SubSec:PhysicsBasedSimulation}

In this study, we focus on the critical aspect of temperature distribution ($T$) in material extrusion-based additive manufacturing processes. Our primary attention is on Fused Filament Fabrication (FFF). These techniques, pivotal in modern manufacturing, involve precise thermal control to ensure product quality. The governing equation for temperature distribution is:
\begin{equation}
\begin{split}
    \rho C_p \frac{\partial T}{\partial t} & = \nabla \cdot \kappa \nabla T \ + \; S(\mathbf{x},t) \; \mathrm{in} \; \Omega_A \\  
    T (z = 0) & = T_0 \; \mathrm{for} \; \Omega_A \times [0,t]\\ 
    \kappa \nabla T \cdot \hat{n} \ & + h (T - T_{\infty}) = 0 \; \mathrm{in} \; \Gamma_A \\ 
\end{split}
\label{eq:PDE}
\end{equation}
Here, $\kappa$ represents the thermal conductivity of the material, $\rho$ its density, and $C_p$ the specific heat capacity. The term $S(\mathbf{x},t)$ represents the heat added when a new voxel is deposited.  A convenient analytical expression is
\[
  S(\mathbf{x},t)=\rho C_p\bigl[T_v-T_\infty\bigr]\,
                  \delta\!\bigl(t-t_A(\mathbf{x})\bigr),
\]
where $t_A(\mathbf{x})$ is the activation time of voxel $\mathbf{x}$, $T_\infty$ is the ambitent temperature of surrounding air, and $T_v$ is the nozzle temperature.  In the time-discrete FEM code the Dirac pulse is implemented by resetting the Gauss-point temperature of the new voxel to $T_v$ at the first step after activation, injecting the sensible heat \(\rho C_p\,(T_v-T_\infty)\) in one time step.  Because $T_v$ is read directly from the printer’s G-code, no empirical calibration of $S$ is required when switching between machines. The boundary of the active printing area is denoted by $\Gamma_A$, which is dynamic as the printing progresses. The simulation assumes a constant temperature, $T_0$, at the base (denoted as $z = 0$), which constitutes a Dirichlet boundary condition. 

\paragraph{Remark on curing or crystallisation heat} Eq.~\eqref{eq:PDE} does not include a separate source term for latent heat released by curing or crystallisation.  This simplification is appropriate for the two materials analysed:

\begin{itemize}
\item \textbf{ABS}: fully amorphous, so no exothermic phase change occurs after deposition.
\item \textbf{PEKK}: semi-crystalline, but differential scanning calorimetry (DSC) shows \emph{no crystallisation peak} when the material is cooled at rates of 20\,$^{\circ}$C\,min$^{-1}$ or faster, which are typical of fused-filament-fabrication cooling \citep[][Fig.\,2]{reber2019polyetherketoneketone}.  Under these conditions the polymer remains essentially amorphous during printing, so an additional latent term is unnecessary.
\end{itemize}

If future work involves a polymer that does crystallise or cure on the timescale of printing, an extra source $S_{\mathrm{lat}}$ can be added to the right-hand side of Eq.\,\eqref{eq:PDE}.

\begin{figure*}[!b]
    \centering
    \includegraphics[width=0.8\linewidth, trim={4.2cm 6cm 3.9cm 5.5cm}, clip]{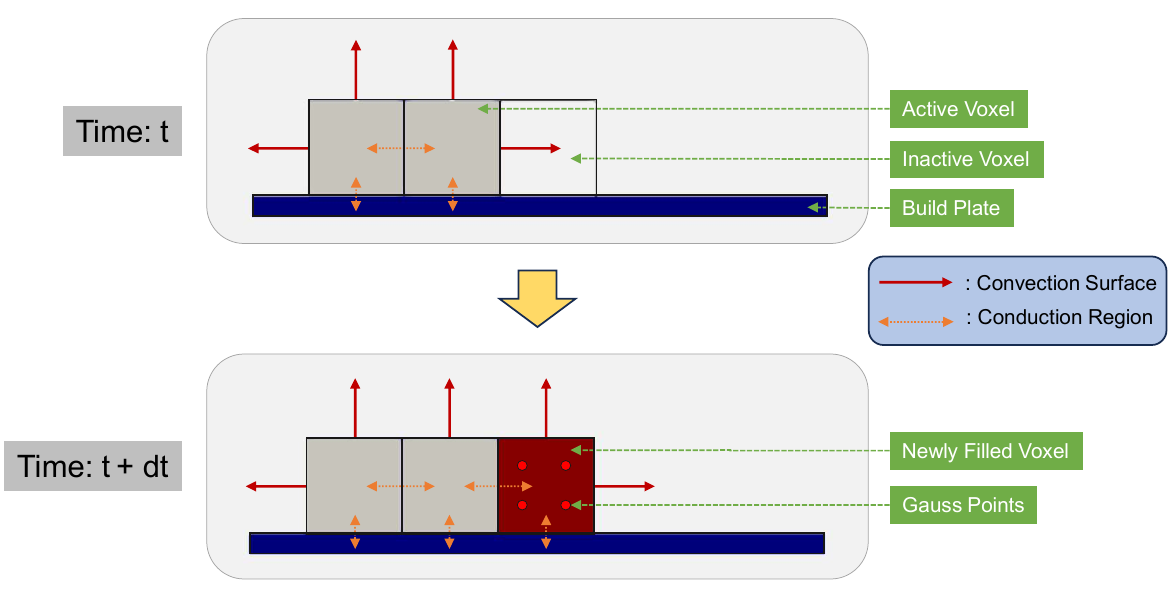}
    \caption{Illustration of the voxel-based deposition process at two consecutive time steps $t$ and $t + \Delta t$. Existing printed voxels (gray) transfer heat through both conduction (orange arrows) and convection (red arrows). At time $t + \Delta t$, a newly filled voxel (red) is added, expanding the active region in which heat is solved. The build plate (blue) is held at a prescribed temperature boundary condition. This schematic highlights how voxel activation, heat conduction, and convection boundaries are dynamically updated throughout the printing process.}
    \label{Fig:voxelInitialization}
\end{figure*}

\paragraph{Non-dimensionalization}
\label{SubSec:NonDimensionalization}
The non-dimensionalized form of the transient heat equation can be obtained by introducing dimensionless variables and scaling factors to the original equation. We assume that the characteristic length of the system is $L$, the characteristic time is $t_c$, and the characteristic temperature difference is $\Delta T_c$. Then, the dimensionless variables are defined as follows. The dimensionless temperature $T'$: $T' = \frac{T - T_{\infty}}{\Delta T_c}$, where $T_{\infty}$ is the ambient temperature. The dimensionless spatial coordinate $x'$: $x' = \frac{x}{L}$. The dimensionless time $t'$: $t' = \frac{t}{t_c}$. Using these dimensionless variables, we can rewrite the transient heat equation as:
\begin{equation}
    \frac{\partial T'}{\partial t'} =\nabla' \cdot  \alpha \nabla' T'
\end{equation}
where $\alpha = \frac{\kappa T_c}{\rho c_p L^2}$ is the dimensionless thermal diffusivity. The non-dimensionalized equation is now independent of the specific values of the characteristic length, time, and temperature difference; it can be used to analyze heat transfer in any system with the same geometry and material properties.

Our simulation methodologically accounts for both conductive and convective heat transfer.  While conduction is modelled through the material to the temperature-controlled bed, convection is represented by the Robin boundary condition applied along~$\Gamma_A$.  We discretise time with the \textbf{BDF2} (backward-difference, second-order) scheme and employ a \emph{single fixed} time step $\Delta t$ throughout each simulation; no adaptive time stepping is used. \figref{Fig:voxelInitialization} illustrates the voxel grid and adaptive octree hierarchy, highlighting the conduction region (printed material) and convection surfaces (outer boundary).

\begin{algorithm}[t!]
\footnotesize
    \caption{\textsc{Simulation Algorithm}}
    \label{alg:physicsSimulation}
    \begin{algorithmic}[1]
\Require $V[]$ (list of voxel in a given order of printing), $V_G$ (Voxel grid),$O_L$ (Octree level for voxel)
\Ensure $T(\Vec{x})$
\State $O_G \leftarrow$ Construct ($l = l_b$) \Comment{Construct octree at base level $l_b \leq O_L$}
\For{ $i \in len(V)$ } \Comment{Loop through the voxel}
\State $v \leftarrow$ V[i]
\State $V_B \leftarrow$ \textsc{Update\_$V_B$($V_B, v, V_G$)} \Comment{Update voxel grid state}

\If{$v \in new\_layer$} \Comment{Check if the voxel is at new layer}

\State $O_G \leftarrow$ \textsc{OctreeRefine($V_G,v,O_G,O_L)$} \Comment{Refine octree for $v$} (\algoref{alg:octreeRefine})
\State $O_G \leftarrow$ \textsc{OctreeCoarsen($V_G,V_B,v,O_G,O_{BLvl})$} \Comment{Coarsen octree for $v$} (\algoref{alg:octreeCoarsen})

\EndIf

\State Reclassify $\Omega_A$ \Comment{Update \Active{}, \textsc{In} and \textsc{Boundary} elements set}
\State $t \leftarrow 0$
\For{$t$ $<$ $T_n$} \Comment{Solve \eqnref{eq:PDE} for fixed number of timestep}
\For{$e \in \Omega_A$} \Comment{Loop through \Active{} elements}
\State perform matrix and vector assembly \Comment{Fill for both volume and surface integrals}
\EndFor
\State Apply Dirichlet Boundary condition on plate and $\Omega_I$
\State Solve system of equation for $T$
\State $t \leftarrow t + \Delta t$ \Comment{Increment time}
\EndFor
\EndFor{}
\item[]
\Return T       \Comment{return the final temperature distribution}
    \end{algorithmic}
\end{algorithm}

\algoref{alg:physicsSimulation} outlines the key steps for simulating the temperature distribution during the additive manufacturing process. The algorithm begins with a queue of voxels arranged in the order of printing. Initially, we create an octree at a base level, which is typically coarser than the required voxel level. The process commences with the removal of a voxel $v$ from the start of the queue. We update $V_B$ to reflect this new addition and to identify new boundary surfaces. If a change in the layer height of printing is observed, we reconstruct the octree $O_G$. This reconstruction is based on the refinement and coarsening strategies outlined in ~\algoref{alg:octreeRefine} and ~\algoref{alg:octreeCoarsen}. Subsequently, the element corresponding to the voxel $v$ is added to the domain $\Omega_A$. With the updated octree and appropriately classified elements, we proceed to solve the heat equation up to $t = T_n$. Here, $T_n$ represents the interval before the addition of the next voxel, which we assume to be constant for each voxel. This process is repeated iteratively for each voxel in the queue. Upon completion, when all voxels are processed, the algorithm outputs the final temperature distribution across the printed object.

\section{Results}
\label{Sec:Results}

In this section, we present a comprehensive set of numerical and experimental results that showcase the capabilities and versatility of our proposed framework for real-time geometric and thermal simulations of extrusion-based additive manufacturing. We organize our findings into four key areas. First, we validate our thermal simulation against experimental data for single-filament wall prints of ABS and PEKK, demonstrating the accuracy of the model in predicting temperature evolution across different materials. Next, we apply our framework to a range of complex geometries—specifically the Stanford Bunny, 3D Benchy, and Moai head—at multiple voxel resolutions to highlight its scalability and ability to capture intricate geometric details. In the third part, we showcase our ability to model infill sparsity and the effect it has on temperature distribution, providing insight into how varying internal structures impact heat transfer. Finally, we present photographs of physical 3D prints of these complex geometries, printed using the \vbv{} approach at the resolution of our simulation, offering a visual and dimensional comparison between the simulated and actual printed parts. We show how our proposed approach can efficiently and accurately handle these diverse scenarios. We also highlight the computational cost and discuss possible limitations and future directions.

\subsection{Comparison against Reported Experimental and Numerical Data for Single Filament Wall}
\label{SubSec:SingleFilamentWall}

\begin{table}[!b]
    \centering
    \setlength{\extrarowheight}{3pt}
    \caption{Material properties and 3D-printer parameters used for the single filament wall simulations.}
    \label{Tab:SingleFilamentWall_mat_process_params}
    \begin{tabular}{|l|r|r|}
        \hline
        \textbf{Property}   & \textbf{ABS} & \textbf{PEKK}\\
        \hline
        \hline
        Density $(\mathrm{kg/m^3})$        & 1050  & 1140\\
        \hline
        Thermal conductivity $(\mathrm{W/m\cdot K})$ & 0.2   & 0.5\\
        \hline
        Specific heat capacity $(\mathrm{J/kg\cdot K})$  & 2100  & 2200\\
        \hline
        Thermal diffusivity $(\mathrm{m^2/s})$        & $9.9\times10^{-8}$  & $1.99\times10^{-7}$\\
        \hline
        Convection coefficient $(\mathrm{W/m^2\cdot K})$ & 30\,\textsuperscript{a} & 30\,\textsuperscript{a}\\
        \hline
        Nozzle temperature $(\mathrm{K})$         & 528.15 & 629.15\\
        \hline
        Print bed temperature $(\mathrm{K})$      & 373.15 & 433.15\\
        \hline 
        Chamber temperature $(\mathrm{K})$        & 368.15 & 412.15\\
        \hline
        Print speed $(\mathrm{mm/s})$             & 6.741  & 6.12\\
        \hline
    \end{tabular}
    \vspace{1ex}\par
    \raggedright\small
    \textsuperscript{a} Adopted directly from \citet{lepoivre2020heat}.
\end{table}

\begin{table}[!b]
    \centering
    \setlength{\extrarowheight}{3pt}
    \caption{Geometry and voxel discretization for the single filament wall.}
    \label{Tab:SingleFilamentWall_GeoInfo}
    \begin{tabular}{|l|c|c|}
        \hline
        \textbf{–}                  & \textbf{ABS}      & \textbf{PEKK}     \\
        \hline
        \hline
        Wall dimensions $(\mathrm{mm})$  & $60 \times 1.25 \times 51.2$  & $60 \times 2.2 \times 51.2$  \\
        \hline
        Voxel size $(\mathrm{mm})$      & $5.0 \times 1.25 \times 0.8$  & $5.0 \times 2.2 \times 0.8$  \\
        \hline
        Voxels per wall $(\#)$          & $12 \times 1 \times 63$       & $12 \times 1 \times 63$       \\
        \hline
        Total number of voxels          & 756                           & 756                           \\
        \hline
    \end{tabular}
\end{table}

\begin{table}[!ht]
    \centering
    \setlength{\extrarowheight}{3pt}
    \caption{Key simulation parameters for the single filament wall.}
    \label{Tab:SingleFilamentWall_sim_params}
    \begin{tabular}{|l|r|r|}
        \hline
        \textbf{–}                     & \textbf{ABS}  & \textbf{PEKK} \\
        \hline
        \hline
        Time step size $(\mathrm{s})$         & 0.1854    & 0.2042   \\
        \hline
        Time steps per new voxel $(\#)$       & 4         & 4        \\
        \hline
        Total number of time steps $(\#)$     & 3024      & 3024     \\
        \hline
        Final DOFs $(\#)$                     & 1612      & 1612     \\
        \hline
        Wall time for each run $(\mathrm{min})$ & 4       & 4        \\
        \hline
    \end{tabular}
\end{table}

\begin{figure}[!t]
  \begin{subfigure}{0.48\linewidth}
    \centering
    \begin{tikzpicture}
      \begin{axis}[
        width=0.99\linewidth, height=7cm,
        xlabel={Time (s)}, ylabel={Temperature (K)},
        xmin=0, xmax=60, ymin=340, ymax=560,
        grid=major, grid style={dashed,gray!30},
        legend style={font=\scriptsize,cells={align=left},draw=none,
                      at={(0.97,0.97)}, anchor=north east}]
        \addplot[color=blue, mark=diamond*, line width=1pt, mark size=1pt]
          table[x index=16,y index=17,col sep=comma]{Data/result_abs_5mm.csv};
        \addplot[color=blue!60!white, mark=o, mark size=.9pt]
          table[x index=16,y index=17,col sep=comma]{Data/result_abs_3mm.csv};
        \addplot[color=blue!30!white, mark=x, mark size=.9pt]
          table[x index=16,y index=17,col sep=comma]{Data/result_abs_2p5mm.csv};
        \addplot[color=red, mark=triangle, line width=1pt, mark size=.9pt]
          table[x index=0,y index=2,col sep=comma]{Data/result_abs_5mm.csv};
        \addplot[color=red!60!white, mark=+, mark size=.9pt]
          table[x index=4,y index=6,col sep=comma]{Data/result_abs_5mm.csv};
        \addplot[color=red!30!white, mark=pentagon*, mark size=.9pt]
          table[x index=8,y index=10,col sep=comma]{Data/result_abs_5mm.csv};
        \legend{
          Sim 5mm, Sim 3mm, Sim 2.5mm,
          Exp \citep{lepoivre2020heat},
          Sim \citep{lepoivre2020heat},
          Lin Sim \citep{nagaraj2023novel}}
      \end{axis}
    \end{tikzpicture}
    \caption{ABS}
    \label{Fig:simToExpABS}
  \end{subfigure}%
  \hfill
  \begin{subfigure}{0.48\linewidth}
    \centering
    \begin{tikzpicture}
      \begin{axis}[
        width=0.99\linewidth, height=7cm,
        xlabel={Time (s)}, ylabel={Temperature (K)},
        xmin=0, xmax=60, ymin=440, ymax=660,
        grid=major, grid style={dashed,gray!30},
        legend style={font=\scriptsize,draw=none,
                      at={(0.97,0.97)}, anchor=north east}]
        \addplot[color=blue, mark=diamond*, line width=1pt, mark size=1pt]
          table[x index=16,y index=17,col sep=comma]{Data/result_pekk_5mm.csv};
        \addplot[color=blue!60!white, mark=o, mark size=.9pt]
          table[x index=16,y index=17,col sep=comma]{Data/result_pekk_3mm.csv};
        \addplot[color=blue!30!white, mark=x, mark size=.9pt]
          table[x index=16,y index=17,col sep=comma]{Data/result_pekk_2p5mm.csv};
        \addplot[color=red, mark=triangle, line width=1pt, mark size=.9pt]
          table[x index=0,y index=2,col sep=comma]{Data/result_pekk_5mm.csv};
        \addplot[color=red!60!white, mark=+, mark size=.9pt]
          table[x index=4,y index=6,col sep=comma]{Data/result_pekk_5mm.csv};
        \addplot[color=red!30!white, mark=pentagon*, mark size=.9pt]
          table[x index=8,y index=10,col sep=comma]{Data/result_pekk_5mm.csv};
        \legend{%
          Sim 5mm, Sim 3mm, Sim 2.5mm,
          Exp \citep{lepoivre2020heat},
          Sim \citep{lepoivre2020heat},
          Lin Sim \citep{nagaraj2023novel}}
      \end{axis}
    \end{tikzpicture}
    \caption{PEKK}
    \label{Fig:simToExpPEKK}
  \end{subfigure}

  \vspace{1em}
  \centering
  \begin{subfigure}{0.5\linewidth}
    \centering
    \includegraphics[width=\linewidth]{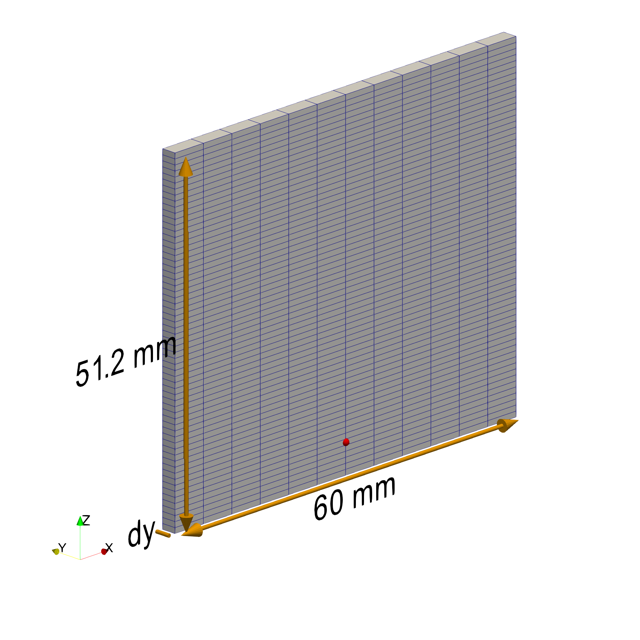}
    \caption{Wall geometry and probe location.}
    \label{Fig:wallGeometryCentered}
  \end{subfigure}

    \caption{\textbf{(a)} Temperature vs. time for ABS: our three voxel resolutions (5 mm, 3 mm, 2.5 mm) versus the experiment and simulation curves from \citet{lepoivre2020heat} and \citet{nagaraj2023novel}. \textbf{(b)} Temperature vs. time fozr PEKK under the same conditions. \textbf{(c)} Single filament wall geometry ($60\,\mathrm{mm}\times 51.2\,\mathrm{mm}$). The thickness is $1.25\,\mathrm{mm}$ for ABS and $2.2\,\mathrm{mm}$ for PEKK; the red dot marks the temperature probe location at $(30\,\mathrm{mm},\,0\,\mathrm{mm},\,4.4\,\mathrm{mm})$. \tabref{Tab:SingleFilamentWall_mat_process_params} - \tabref{Tab:SingleFilamentWall_sim_params} provide material, process, geometry, and numerical details.}
  \label{Fig:SingleFilamentWallTemperature}
\end{figure}

To establish a literature benchmark, we compare our transient‐thermal predictions against both the experimental measurements and numerical simulations reported by \citet{lepoivre2020heat} (experiment + simulation) and \citet{nagaraj2023novel} (linear‐element simulation). All convection coefficients (\(h=30\,\mathrm{W/m^2\,K}\)) are adopted directly from \citet{lepoivre2020heat}, Section 2.3, without independent calibration. While this choice ensures consistency, it also means that any agreement is partly due to the same parameter selection.

\tabref{Tab:SingleFilamentWall_mat_process_params} through \tabref{Tab:SingleFilamentWall_sim_params} summarize all of the key inputs for our single‐filament‐wall simulations. \tabref{Tab:SingleFilamentWall_mat_process_params} gives the material properties and 3D-printer settings. \tabref{Tab:SingleFilamentWall_GeoInfo} shows the wall geometry and voxel discretization: each wall measures 60 mm × 51.2 mm, with a thickness of 1.25 mm (ABS) or 2.2 mm (PEKK), voxelized into a 12 × 1 × 63 grid (756 total voxels). Finally, \tabref{Tab:SingleFilamentWall_sim_params} lists the numerical parameters: a fixed time step of approximately 0.19 s (four steps per newly activated voxel), 3024 total time steps, and a final degree-of-freedom count of 1612. Each simulation required about 4 minutes on our standard workstation.

\figref{Fig:SingleFilamentWallTemperature} overlays our voxel-FEM predictions (blue) on the experimental thermocouple trace and finite-element results of \citet{lepoivre2020heat} (red) as well as the linear-element simulation of \citet{nagaraj2023novel} (light-red). Both ABS and PEKK curves exhibit multiple temperature rises (“bumps”) the first of which occurs at \(t\!\approx\!9\;\mathrm{s}\) for ABS and \(t\!\approx\!10\;\mathrm{s}\) for PEKK.  These bumps are caused by the addition of new voxels in above layers, which reheats the previously solidified material. The first bump is the largest, followed by smaller bumps. The magnitude of these bumps decreases with increasing distance from the probe layer.

For ABS (\figref{Fig:simToExpABS}) our peak temperature and the subsequent cooling slope follow the experimental data within \(\pm2\;\mathrm{K}\).  Agreement with the two reference simulations is slightly looser—within \(\pm15\;\mathrm{K}\)—but the overall cooling rate is virtually identical, confirming that the constant-property model combined with \(h=30\;\mathrm{W\,m^{-2}K^{-1}}\) reproduces the transient history measured by
Lepoivre et al.

We can infer the following two insights from \figref{Fig:simToExpPEKK}. Comparing our simulation with the experimental data of \citet{lepoivre2020heat}, the curves for \(t<7~\mathrm{s}\)  coincide (difference \(<3~\mathrm{K}\)). Between \(7\) and \(15~\mathrm{s}\) the experiment is as much as \(25~\mathrm{K}\) higher than our prediction; \citet{lepoivre2020heat}report a comparable 25 °C gap in this interval and ascribe it to imperfect chamber-temperature control and pyrometer calibration \citep[Sec. 3.3]{lepoivre2020heat}. After two additional layers are deposited (\(t\gtrsim25~\mathrm{s}\)) the difference drops below \(8~\mathrm{K}\) and remains small for the rest of the record. Comparing our simulation with the FE results of \citet{lepoivre2020heat} and \citet{nagaraj2023novel}, we find that up to \(t\approx10~\mathrm{s}\) all three simulations show the same cooling slope, although the Lepoivre/Nagaraj peaks are about \(15~\mathrm{K}\) higher than ours. From \(t=12\) to \(30~\mathrm{s}\) the three simulation curves cluster within a few Kelvin of each other; beyond \(30~\mathrm{s}\) they diverge, with our prediction staying closest to the experimental trace, whereas the two reference simulations drift below it.

\paragraph{Reasons for the temperature differences} All observations above fall within the discrepancy envelope documented by \citet{lepoivre2020heat}for PEKK: up to 25 °C in the first instants, less than 10 °C from 3 s to 50 s, and again up to 25 °C after 50 s \citep[Sec. 3.3]{lepoivre2020heat}. They identify four contributing factors, which likewise explain the gaps between our curve and theirs:

\begin{itemize}[leftmargin=*]
\item \textit{Heating-chamber control} – temperature drifts can generate the 25 °C error band seen after 50 s.
\item \textit{Pyrometer calibration} – inaccurate emissivity for PEKK causes systematic under- or over-reading, particularly in the first few seconds.
\item \textit{Temperature-dependent material properties and convective coefficient} – using constant \(k\), \(c_p\) and \(h\) (as in all three simulations) leaves out known \(T\)-dependence.
\item \textit{Thermal-contact resistance} – \citet{lepoivre2020heat}note that its value was not measured precisely, adding further uncertainty.
\end{itemize}

Overall, the ABS history matches within experimental error, and the PEKK profile is within the ±25 °C band reported by Lepoivre et al., validating the ability of the proposed framework to capture short-time reheating events and long-time cooling trends with a single, literature-derived convection coefficient.

\subsection{Complex Geometry Simulations}
\label{SubSec:ComplexGeometrySimulations}

We now demonstrate the capability of our framework to handle more intricate shapes, including the well-known Stanford Bunny, 3D Benchy, and Moai head models---common benchmark geometries in computer graphics and 3D printing tests. By varying the voxel resolution, we show how our framework adapts to capture geometric fidelity and thermal effects at different scales.

\paragraph{Geometry and Voxel Information}
\label{SubSubSec:ComplexGeoVoxelInfo}
\tabref{Tab:ComplexGeo_GeoInfo} outlines the bounding box, voxel dimension of the geometry, voxel size, and print voxel counts for these complex geometries. We consider voxel resolutions of $32^3$, $64^3$, and $128^3$ for each geometry to showcase a range of fidelity levels. In each case, we position the geometry such that the $z$-axis corresponds to the building direction (layer stacking).

\begin{table}[!b]
    \centering
    \setlength{\extrarowheight}{3pt}
    \caption{Geometry and voxel information for complex geometry simulations (Stanford Bunny, 3D Benchy, and Moai).}
    \label{Tab:ComplexGeo_GeoInfo}
    \begin{tabular}{|c|c|c|c|r|}
        \hline
        \textbf{Geometry} & \textbf{Bounding box} & \textbf{Voxel dim} & \textbf{Voxel size} & \textbf{Print voxels} \\
         & $(mm)$ & $(\#, (x, y, z))$ & $(mm)$ & $(\#)$ \\
        \hline
        \hline
        \multirow{3}{*}{Bunny}   & $28.2 \times 36.2 \times 35.9$    & $24 \times 32 \times 32$      & $1.18 \times 1.13 \times 1.12$ & $7,871$ \\
        \cline{2-5}
            & $28.2 \times 36.2 \times 35.9$    & $52 \times 64 \times 64$      & $0.54 \times 0.57 \times 0.56$ & $60,008$ \\
        \cline{2-5}
            & $28.2 \times 36.2 \times 35.9$    & $100 \times 128 \times 128$   & $0.28 \times 0.28 \times 0.28$ & $464,158$ \\
        \hline
        \multirow{3}{*}{Benchy}  & $19.1 \times 36.5 \times 29.3$    & $16 \times 32 \times 28$      & $1.19 \times 1.14 \times 1.05$ & $4,070$ \\
        \cline{2-5}
            & $19.1 \times 36.5 \times 29.3$    & $36 \times 64 \times 52$      & $0.53 \times 0.57 \times 0.56$ & $26,479$ \\
        \cline{2-5}
            & $19.1 \times 36.5 \times 29.3$    & $68 \times 128 \times 104$    & $0.28 \times 0.29 \times 0.28$ & $175,481$ \\
        \hline
        \multirow{3}{*}{Moai}   & $27.0 \times 20.3 \times 35.5$    & $24 \times 20 \times 32$      & $1.12 \times 1.02 \times 1.11$ & $3,943$ \\
        \cline{2-5}
            & $27.0 \times 20.3 \times 35.5$    & $48 \times 36 \times 64$      & $0.56 \times 0.56 \times 0.55$ & $23,631$ \\
        \cline{2-5}
            & $27.0 \times 20.3 \times 35.5$    & $100 \times 76 \times 128$    & $0.27 \times 0.27 \times 0.28$ & $187,197$ \\
        \hline
    \end{tabular}
\end{table}

\paragraph{Thermal Simulations at Multiple Resolutions}
\label{SubSubSec:ComplexGeoThermalSims}

\begin{figure*}[t!]
    \centering
    \begin{tikzpicture}[x=3.3cm, y=-3.0cm]
        \node[anchor=north west] (bunny_32_30) at (0,0)
            {\includegraphics[width=0.16\linewidth]{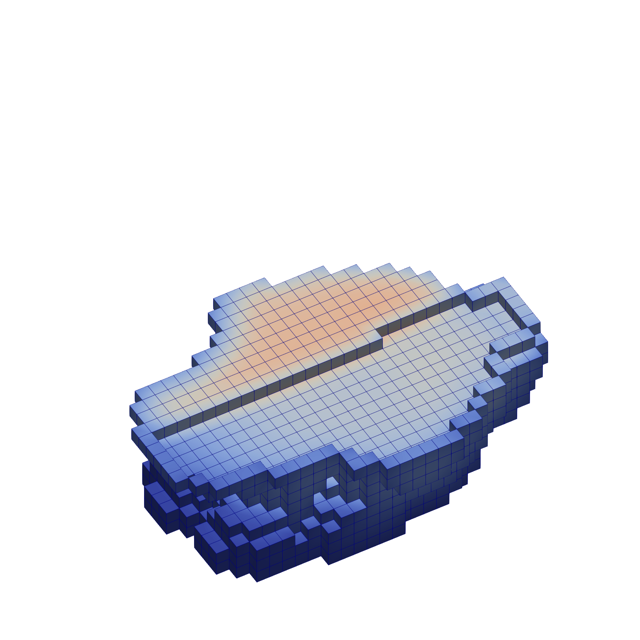}};
        \node[anchor=north west] (bunny_32_60) at (1,0)
            {\includegraphics[width=0.16\linewidth]{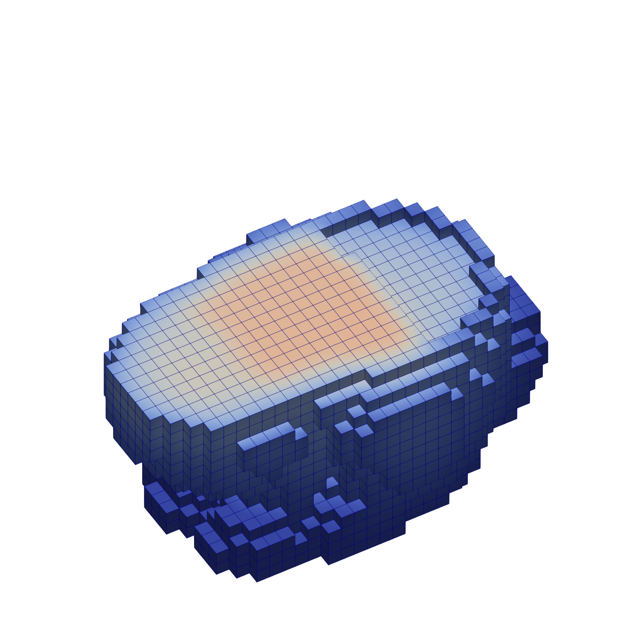}};
        \node[anchor=north west] (bunny_32_100) at (2,0)
            {\includegraphics[width=0.16\linewidth]{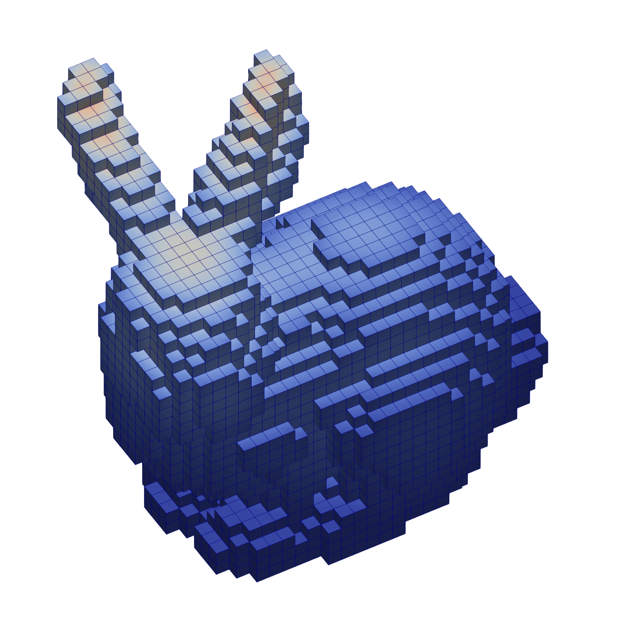}};

        \node[anchor=north west] (bunny_64_30) at (0,.8)
            {\includegraphics[width=0.16\linewidth]{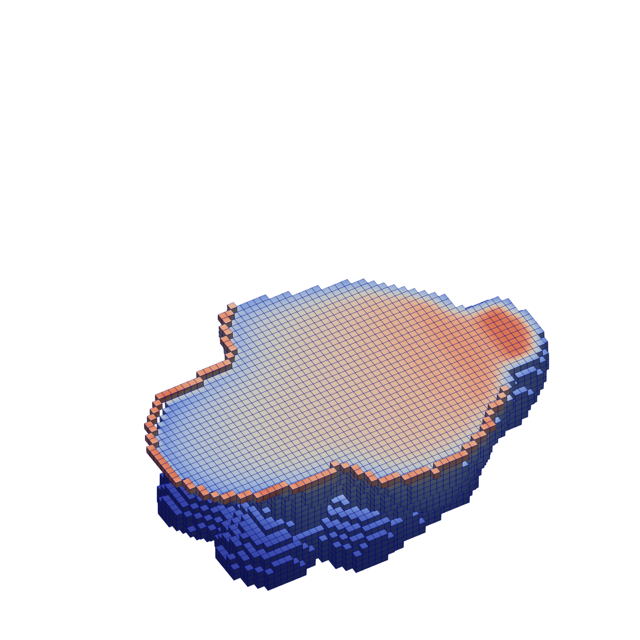}};
        \node[anchor=north west] (bunny_64_60) at (1,.8)
            {\includegraphics[width=0.16\linewidth]{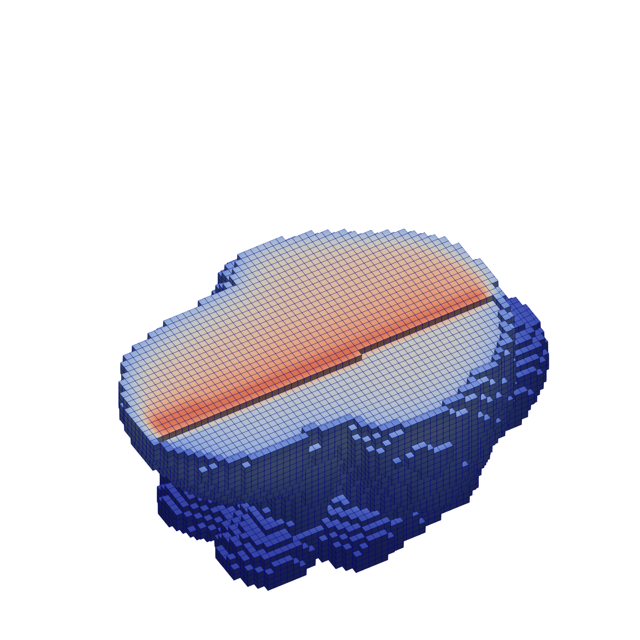}};
        \node[anchor=north west] (bunny_64_100) at (2,.8)
            {\includegraphics[width=0.16\linewidth]{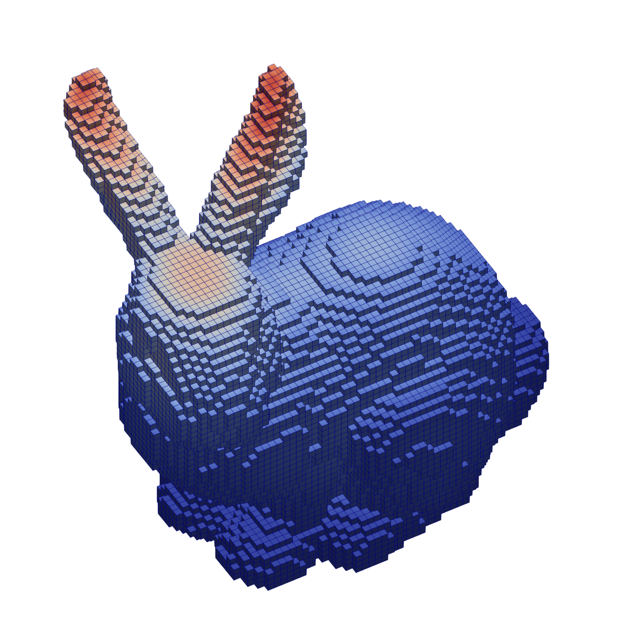}};

        \node[anchor=north west] (bunny_128_30) at (0,1.6)
            {\includegraphics[width=0.16\linewidth]{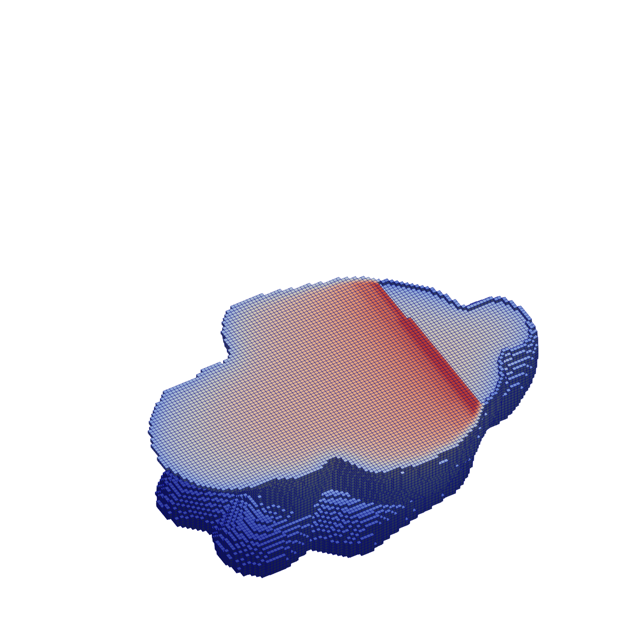}};
        \node[anchor=north west] (bunny_128_60) at (1,1.6)
            {\includegraphics[width=0.16\linewidth]{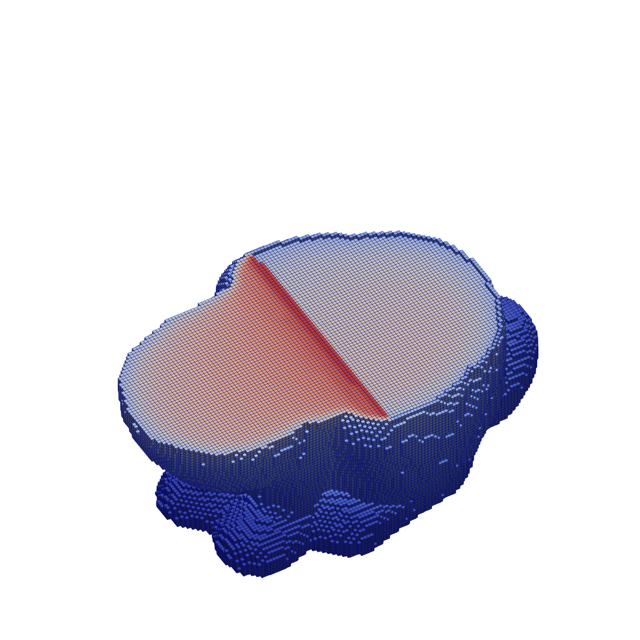}};
        \node[anchor=north west] (bunny_128_100) at (2,1.6)
            {\includegraphics[width=0.16\linewidth]{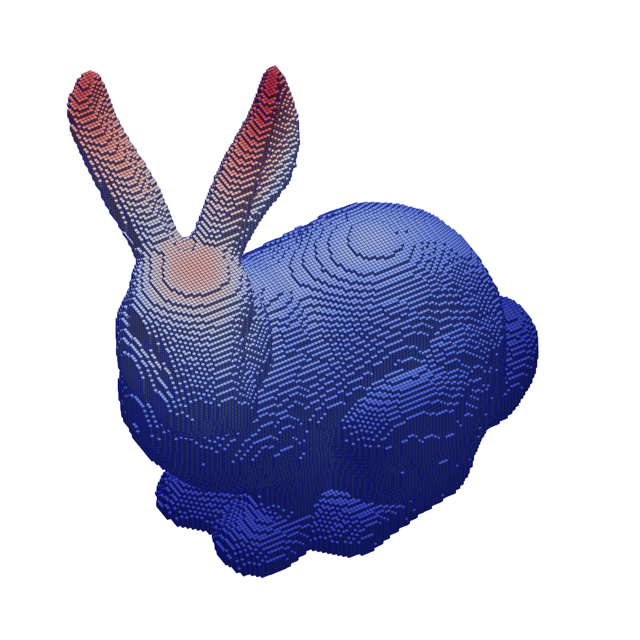}};
        
        \node[above=0.1cm] at (bunny_32_30.north)   {30\%};
        \node[above=0.1cm] at (bunny_32_60.north)   {60\%};
        \node[above=0.1cm] at (bunny_32_100.north)  {100\%};

        \node[left=0.1cm] at (bunny_32_30.west)     {32$^3$};
        \node[left=0.1cm] at (bunny_64_30.west)     {64$^3$};
        \node[left=0.1cm] at (bunny_128_30.west)    {128$^3$};

        \node[rotate=90, below=0.1cm] at (bunny_64_100.east) {\includegraphics[width=6cm, trim={9cm 16cm 8cm 12cm}, clip]{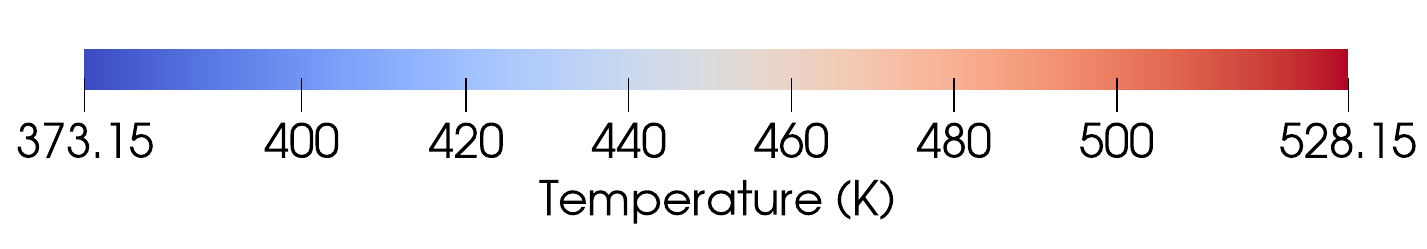}};
    \end{tikzpicture}
    \caption{Stanford Bunny at three voxel resolutions (32, 64, 128) and at three build completion times (30\%, 60\%, 100\%).}
    \label{Fig:complex_geo_bunny}
\end{figure*}
\begin{figure*}[t!]
    \centering
    \begin{tikzpicture}[x=3.3cm, y=-3.0cm]
        \node[anchor=north west] (benchy_32_30) at (0,0)
            {\includegraphics[width=0.16\linewidth]{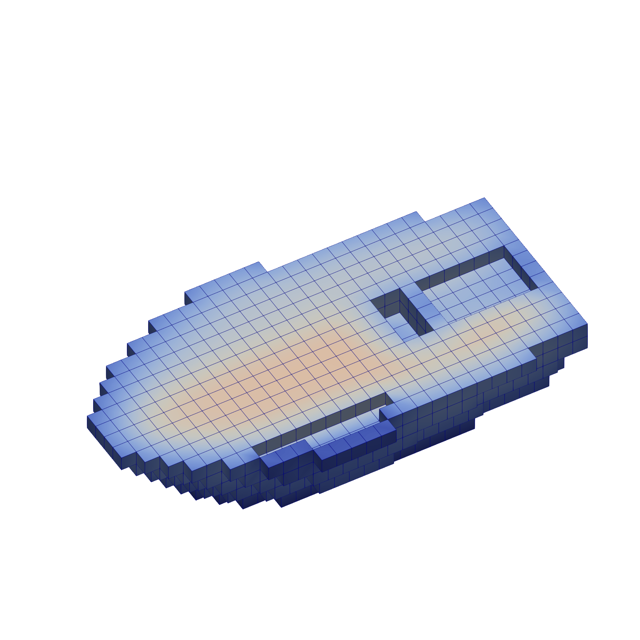}};
        \node[anchor=north west] (benchy_32_60) at (1,0)
            {\includegraphics[width=0.16\linewidth]{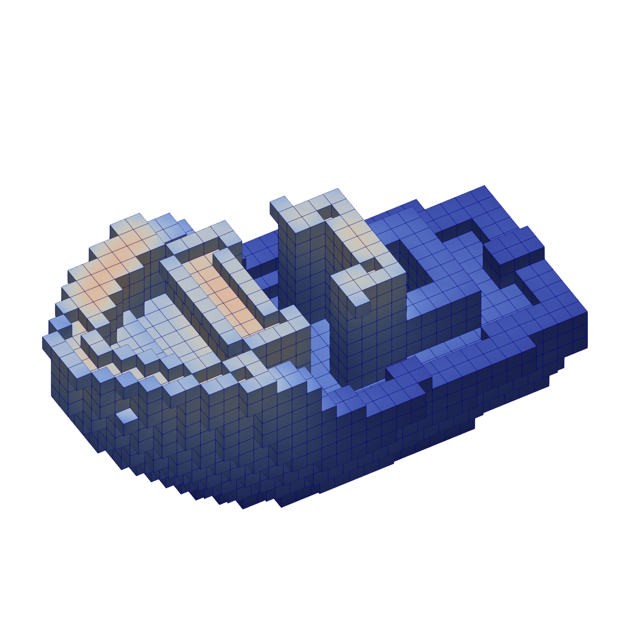}};
        \node[anchor=north west] (benchy_32_100) at (2,0)
            {\includegraphics[width=0.16\linewidth]{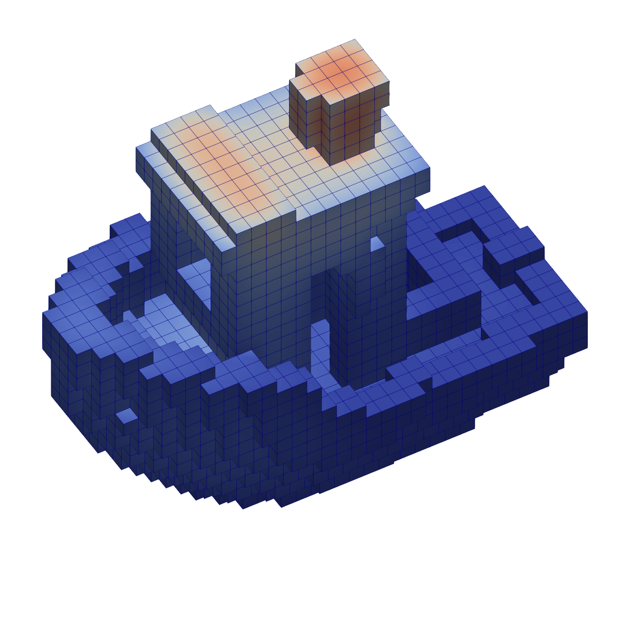}};

        \node[anchor=north west] (benchy_64_30) at (0,.8)
            {\includegraphics[width=0.16\linewidth]{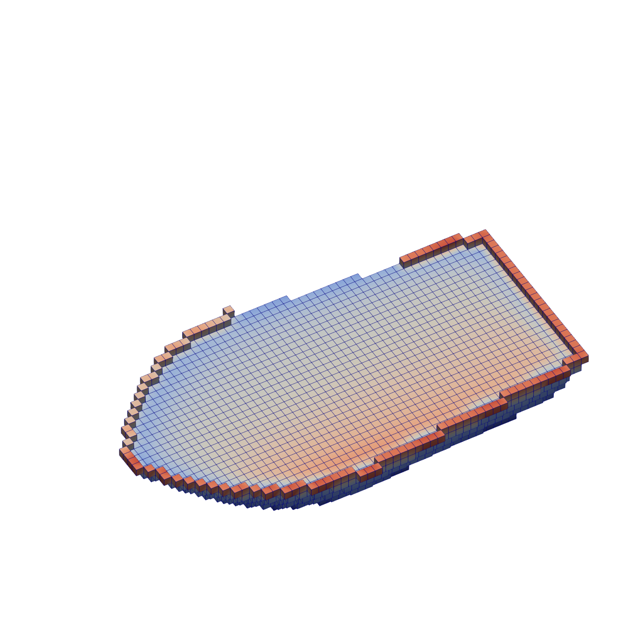}};
        \node[anchor=north west] (benchy_64_60) at (1,.8)
            {\includegraphics[width=0.16\linewidth]{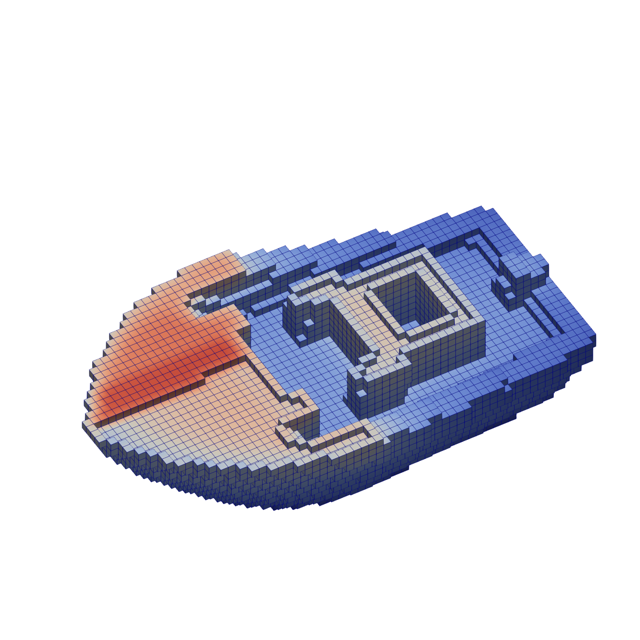}};
        \node[anchor=north west] (benchy_64_100) at (2,.8)
            {\includegraphics[width=0.16\linewidth]{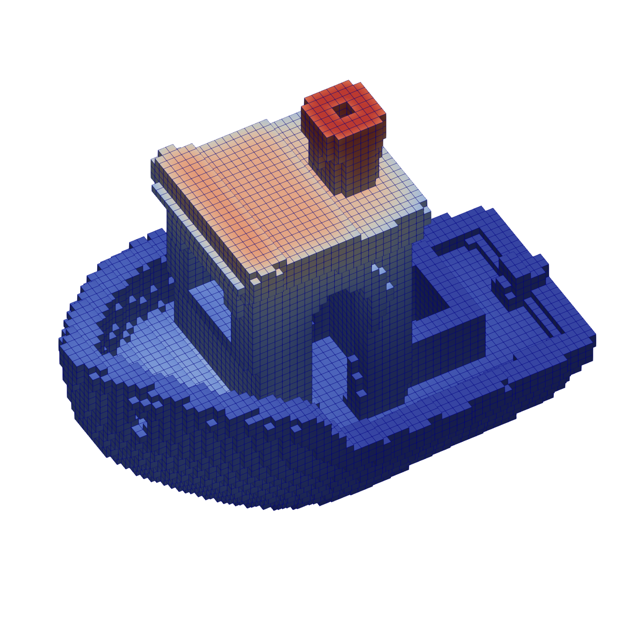}};

        \node[anchor=north west] (benchy_128_30) at (0,1.6)
            {\includegraphics[width=0.16\linewidth]{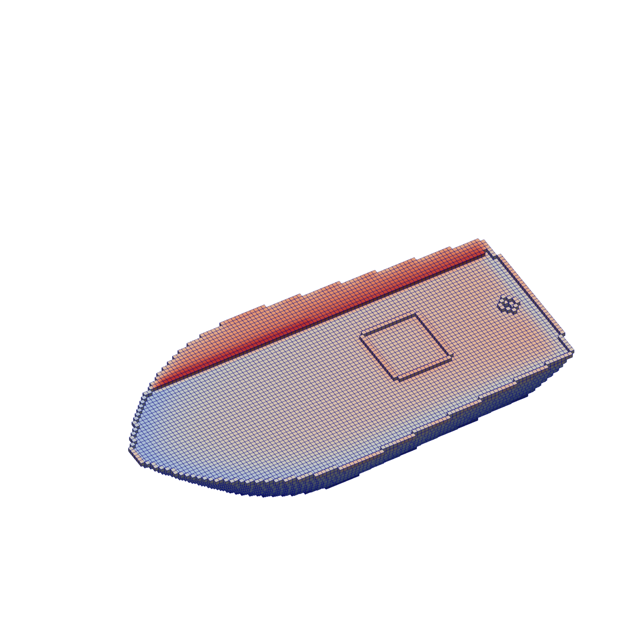}};
        \node[anchor=north west] (benchy_128_60) at (1,1.6)
            {\includegraphics[width=0.16\linewidth]{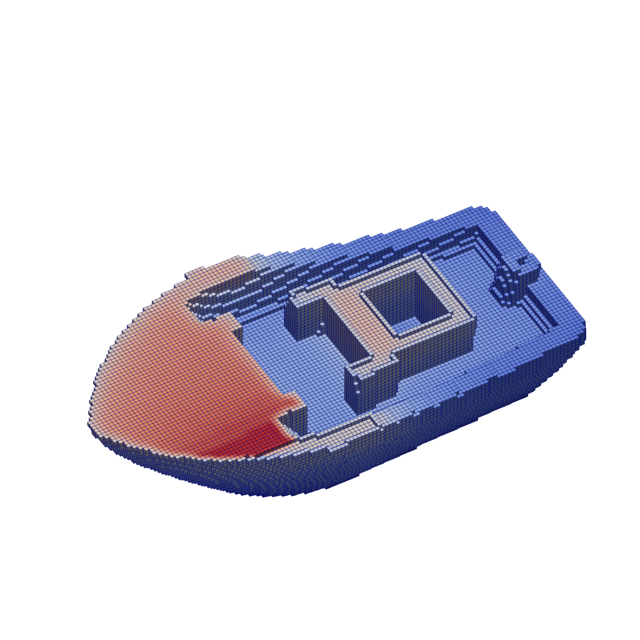}};
        \node[anchor=north west] (benchy_128_100) at (2,1.6)
            {\includegraphics[width=0.16\linewidth]{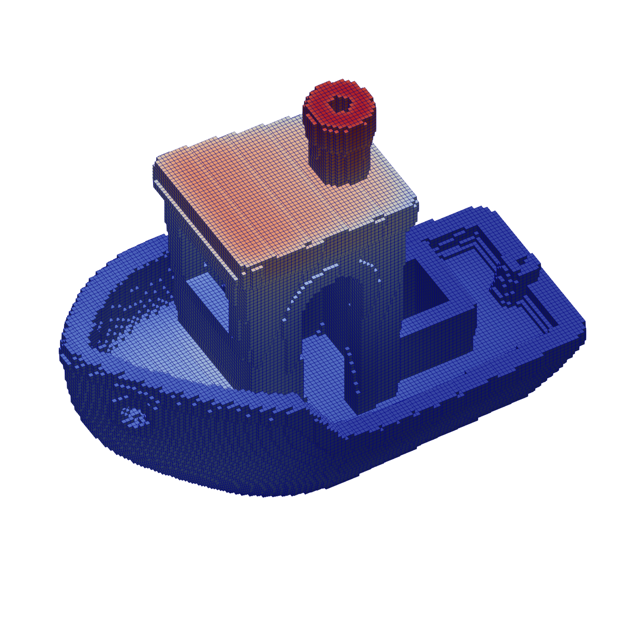}};

        \node[above=0.1cm] at (benchy_32_30.north)  {30\%}; 
        \node[above=0.1cm] at (benchy_32_60.north)  {60\%}; 
        \node[above=0.1cm] at (benchy_32_100.north) {100\%};

        \node[left=0.1cm] at (benchy_32_30.west)    {32$^3$};
        \node[left=0.1cm] at (benchy_64_30.west)    {64$^3$};
        \node[left=0.1cm] at (benchy_128_30.west)   {128$^3$};

        \node[rotate=90, below=0.1cm] at (benchy_64_100.east) {\includegraphics[width=6cm, trim={9cm 16cm 8cm 12cm}, clip]{Figures/kelvin_abs_scale.pdf}};
    \end{tikzpicture}
    \caption{Benchy at three voxel resolutions (32, 64, 128) and at three build completion times (30\%, 60\%, 100\%).}
    \label{Fig:complex_geo_benchy}
\end{figure*}

\begin{figure*}[!ht]
    \centering
    \begin{tikzpicture}[x=3.3cm, y=-3.0cm]
        \node[anchor=north west] (moai_32_30) at (0,0)
            {\includegraphics[width=0.16\linewidth]{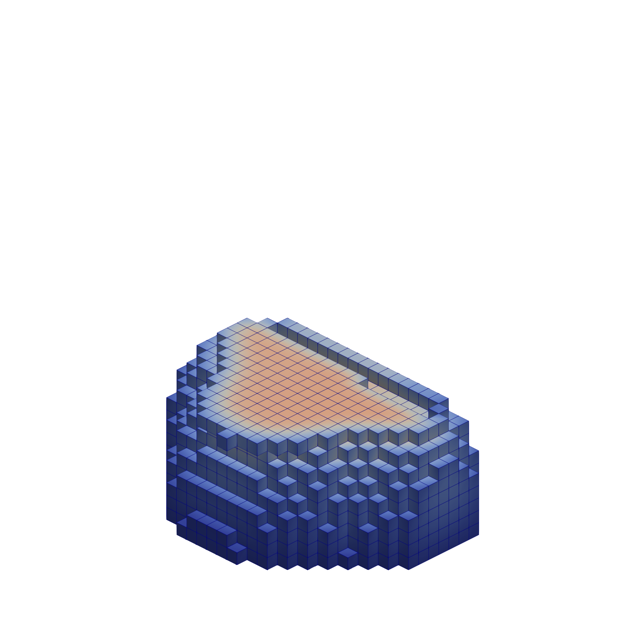}};
        \node[anchor=north west] (moai_32_60) at (1,0)
            {\includegraphics[width=0.16\linewidth]{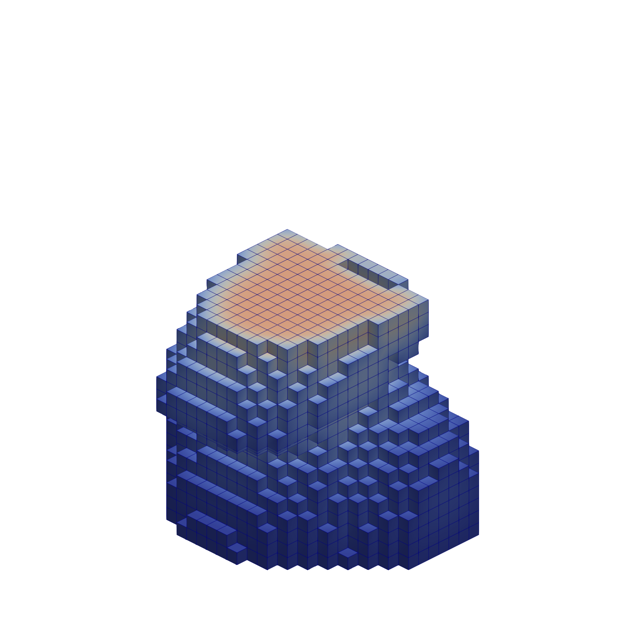}};
        \node[anchor=north west] (moai_32_100) at (2,0)
            {\includegraphics[width=0.16\linewidth]{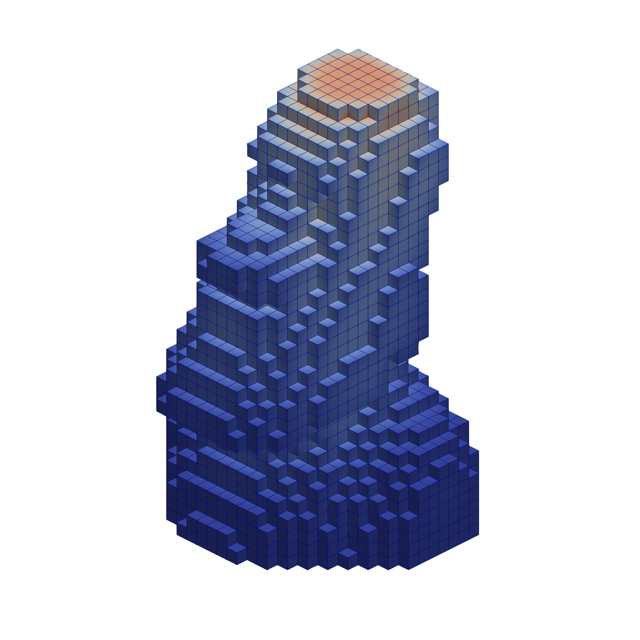}};

        \node[anchor=north west] (moai_64_30) at (0,.8)
            {\includegraphics[width=0.16\linewidth]{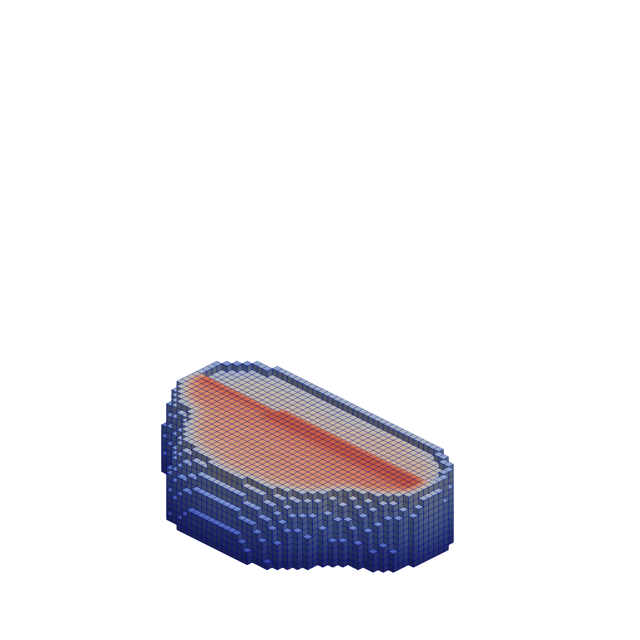}};
        \node[anchor=north west] (moai_64_60) at (1,.8)
            {\includegraphics[width=0.16\linewidth]{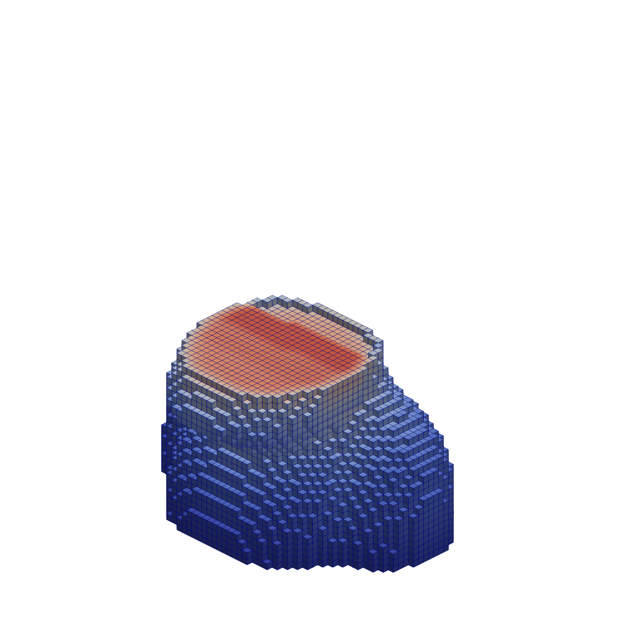}};
        \node[anchor=north west] (moai_64_100) at (2,.8)
            {\includegraphics[width=0.16\linewidth]{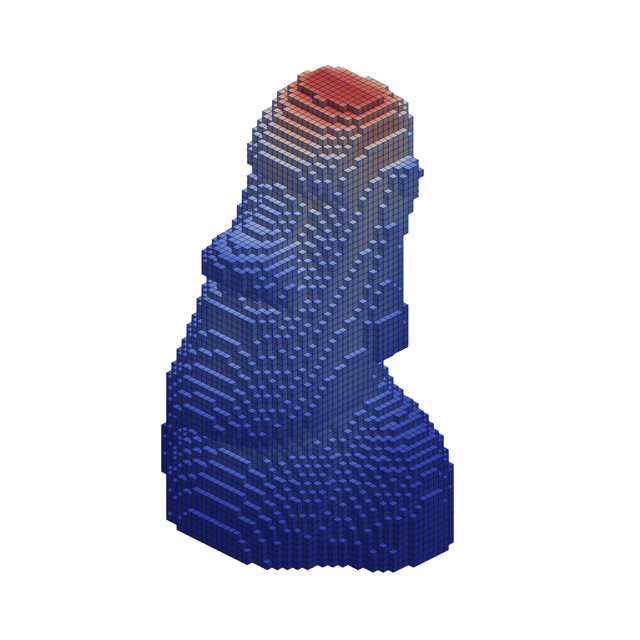}};

        \node[anchor=north west] (moai_128_30) at (0,1.6)
            {\includegraphics[width=0.16\linewidth]{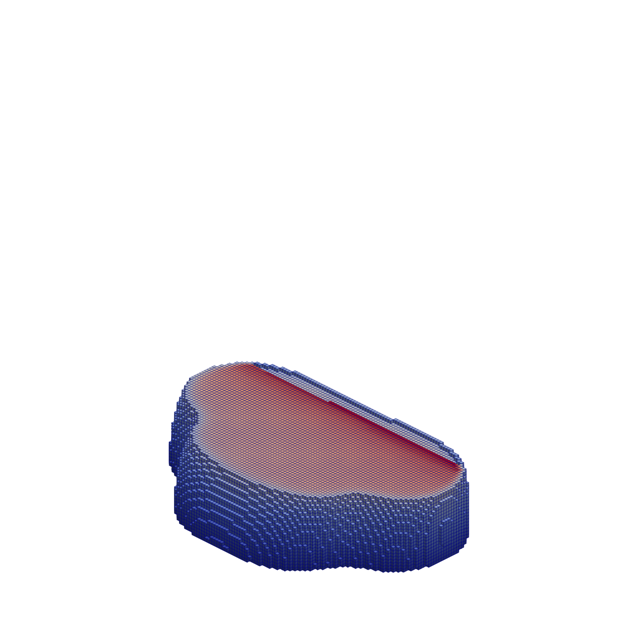}};
        \node[anchor=north west] (moai_128_60) at (1,1.6)
            {\includegraphics[width=0.16\linewidth]{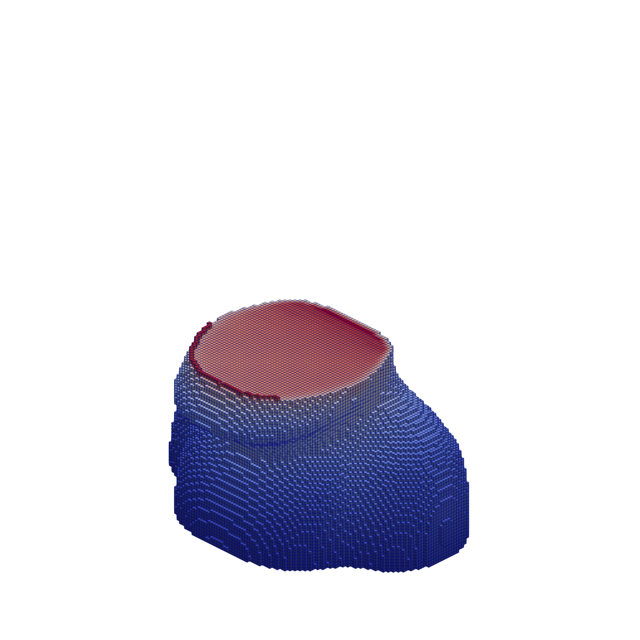}};
        \node[anchor=north west] (moai_128_100) at (2,1.6)
            {\includegraphics[width=0.16\linewidth]{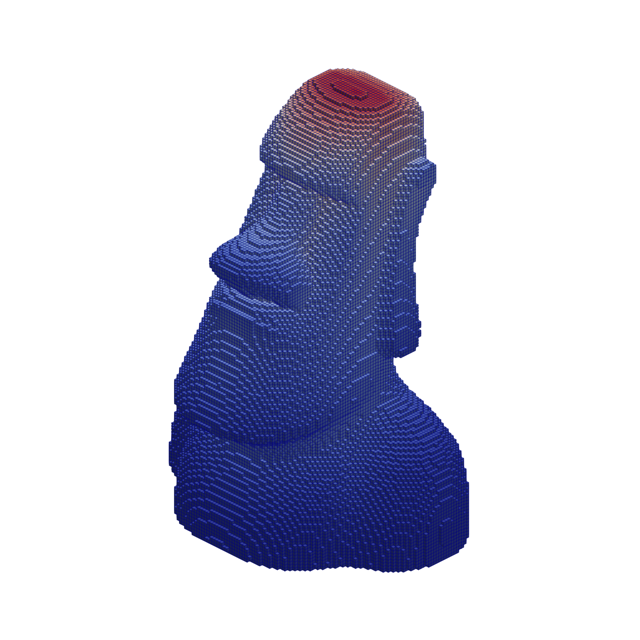}};

        \node[above=0.1cm] at (moai_32_30.north) {30\%};    
        \node[above=0.1cm] at (moai_32_60.north) {60\%};    
        \node[above=0.1cm] at (moai_32_100.north){100\%};   

        \node[left=0.1cm] at (moai_32_30.west)  {32$^3$};
        \node[left=0.1cm] at (moai_64_30.west)  {64$^3$};
        \node[left=0.1cm] at (moai_128_30.west) {128$^3$};


        \node[rotate=90, below=0.1cm] at (moai_64_100.east) {\includegraphics[width=6cm, trim={9cm 16cm 8cm 12cm}, clip]{Figures/kelvin_abs_scale.pdf}};

    \end{tikzpicture}
    \caption{Moai displayed in a 3$\times$3 grid 
    for three resolutions (32, 64, 128) and three completion times (30\%, 60\%, 100\%).}
    \label{Fig:complex_geo_moai}
\end{figure*}

We performed thermal simulations of the Stanford Bunny, 3D Benchy, and Moai models. \figref{Fig:complex_geo_bunny}, \figref{Fig:complex_geo_benchy} and \figref{Fig:complex_geo_moai} shows the temperature evolution at $32^3$, $64^3$ and  $128^3$ resolutions for the Bunny, 3D Benchy and Moai at three snapshots ($30\%$, $60\%$ and $100\%$) of the build. The color scale indicates the temperature in Kelvin with the minimum and maximum values set to print bed and nozzle temperatures, respectively. Each subfigure shows a $3\,\times\,3$ grid for geometry at different resolutions and progress levels during the build. The temperature field evolves from the nozzle temperature to the print bed temperature as the build progresses. The temperature is the highest at the newly deposited voxels and gradually decreases towards the previously printed layers. The cooling effect is due to the heat conduction from the isothermal print bed as well as due to the convection losses to the surrounding air. The color scale is consistent across all resolutions and geometries, allowing for visual comparison of temperature profiles. The simulations are performed using the ABS material properties (\tabref{Tab:SingleFilamentWall_mat_process_params}).

\pagebreak

\paragraph{Simulation Parameters and Computational Cost}
\label{SubSubSec:ComplexGeoSimParams}

\begin{table}[!t]
    \centering
    \setlength{\extrarowheight}{3pt}
    \caption{Representative simulation parameters and scaling behavior for the Bunny, Benchy, and Moai at various resolutions.}
    \label{Tab:ComplexGeo_sim_params}
    \begin{tabular}{|c|c|r|r|r|r|}
        \hline
        \textbf{Geometry} & \textbf{Resolution} & \textbf{Timesteps} & \textbf{Final DOFs} & \textbf{Wall time (s)} & \textbf{Max Processors} \\
        \hline
        \hline
        \multirow{3}{*}{Bunny}   & $32^3$    & $440$     & $140,624$     & $38.76$   & $32$ \\
        \cline{2-6}
            & $64^3$    & $3,280$   & $699,000$     & $492.62$  & $64$ \\
        \cline{2-6}
            & $128^3$   & $21,940$  & $2,680,392$   & $2901.46$ & $256$ \\
        \hline
        \multirow{3}{*}{Benchy}  & $32^3$    & $232$     & $147,120$     & $24.14$   & $32$ \\
        \cline{2-6}
            & $64^3$    & $2,344$   & $619,424$     & $432.3$   & $64$ \\
        \cline{2-6}
            & $128^3$   & $13,520$  & $2,714,888$   & $3334.1$  & $256$ \\
        \hline
        \multirow{3}{*}{Moai}    & $32^3$    & $328$     &  $118,504$    & $39.12$   & $32$ \\
        \cline{2-6}
            & $64^3$    & $2,048$   & $474,104$     & $299.14$  & $64$ \\
        \cline{2-6}
            & $128^3$   & $23,750$  & $2,049,104$   & $2588.81$ & $256$ \\
        \hline
    \end{tabular}
\end{table}

\tabref{Tab:ComplexGeo_sim_params} illustrates the simulation metrics and scalability behavior. As the geometry resolution increases from $32^3$ to $128^3$, the number of time steps and final degrees of freedom (DOFs) also increase significantly. For instance, the Bunny at $128^3$ resolution requires $21,940$ time steps and results in a final DOF count of $2,680,392$. The wall time for simulations also scales with the resolution, with the $128^3$ Bunny simulation taking approximately $2901.46$ seconds on a maximum of $256$ processors. This trend is consistent across all geometries, with the Moai at $128^3$ taking about $2588.81$ seconds and the Benchy at $128^3$ taking around $3334.1$ seconds. Our framework, integrated with \petsc{}, enables efficient parallelization using BiCGS solvers with Additive Schwartz preconditioning. In previous work \citep{saurabh2022scalable}, we demonstrated scalability to $\mathcal{O}(100K)$ processors. This demonstrates the computational efficiency of our framework, as it can handle complex geometries with a high degree of fidelity while maintaining reasonable simulation times. 

\subsection{Geometries with Sparse Infill}
\label{SubSec:SparselyInfillGeometries}

Infill patterns significantly affect both the mechanical and thermal response of a 3D-printed part. To investigate this, we performed simulations with varying infill densities on the Bunny models at a $128^3$ resolution; the geometry details are shown in \tabref{Tab:SparseInfill_sim_details}. We define sparsity by ``skipping'' some fraction of voxels in the interior based on a user-defined pattern (e.g., skipping 3 of every 4 interior rows).

\begin{table}[!b]
    \centering
    \setlength{\extrarowheight}{3pt}
    \caption{Sparse infill simulation details for $128^3$ Stanford Bunny at different \% sparse layers, starting with no sparsity (0\%) and increasing to 70\% sparse layers.}
    \label{Tab:SparseInfill_sim_details}
    \begin{tabular}{|c|r|r|r|r|r|}
        \hline
        \textbf{\% Sparse Layers} & \textbf{Print Voxels} & \textbf{Time Steps} & \textbf{Final DOFs} & \textbf{Wall Time (s)} & \textbf{Processors} \\
        \hline
        \hline
        $0$  & $464,158$ & $21,940$ & $2,680,392$ & $2901.46$ & $256$\\
        \hline
        $30$ & $280,842$ & $14,200$ & $4,470,880$ & $2466.72$ & $256$\\
        \hline
        $50$ & $194,393$ & $9,840$ & $5,108,496$ & $1688.63$ & $256$\\
        \hline
        $70$ & $169,226$ & $8,560$ & $5,232,648$ & $1385.83$ & $256$\\
        \hline
    \end{tabular}
\end{table}

\begin{figure*}[!b]
    \centering
    \begin{tikzpicture}[x=3.3cm, y=-3.0cm]
        \node[anchor=north west] (surf_0p0_bunny) at (0,0)
            {\includegraphics[width=0.16\linewidth]{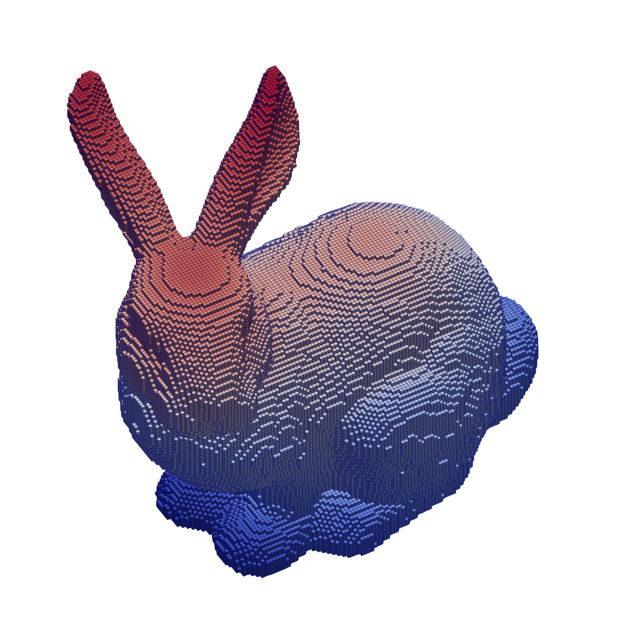}};
        \node[anchor=north west] (surf_0p3_bunny) at (1,0)
            {\includegraphics[width=0.16\linewidth]{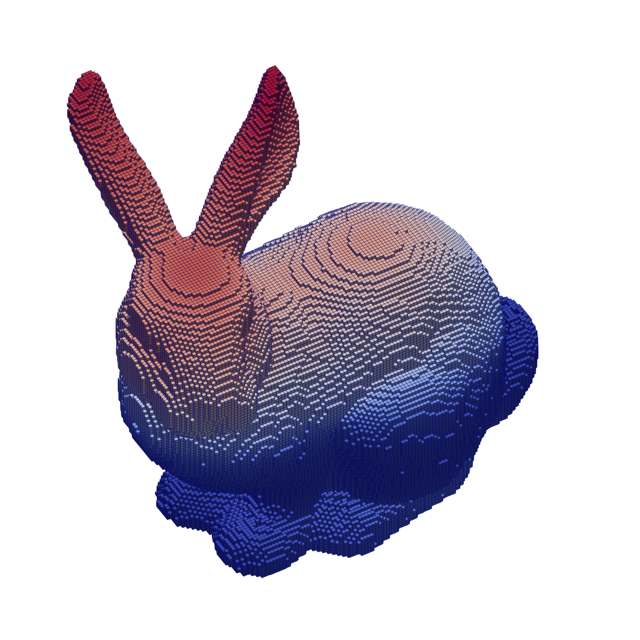}};
        \node[anchor=north west] (surf_0p5_bunny) at (2,0)
            {\includegraphics[width=0.16\linewidth]{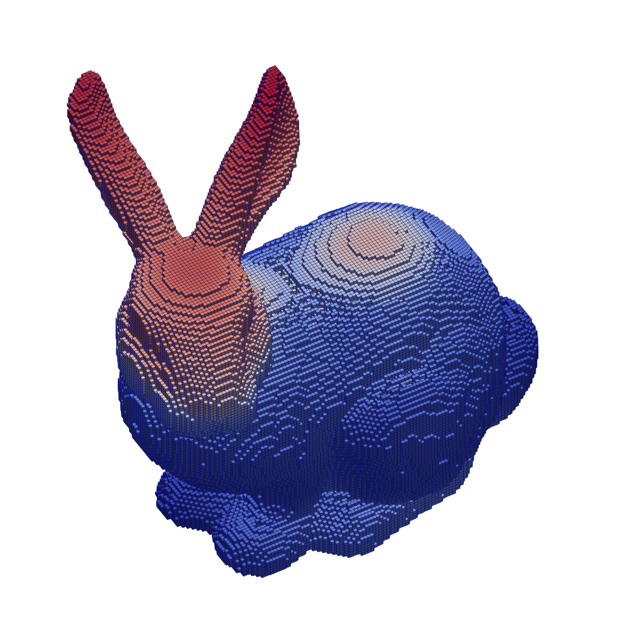}};
        \node[anchor=north west] (surf_0p7_bunny) at (3,0)
            {\includegraphics[width=0.16\linewidth]{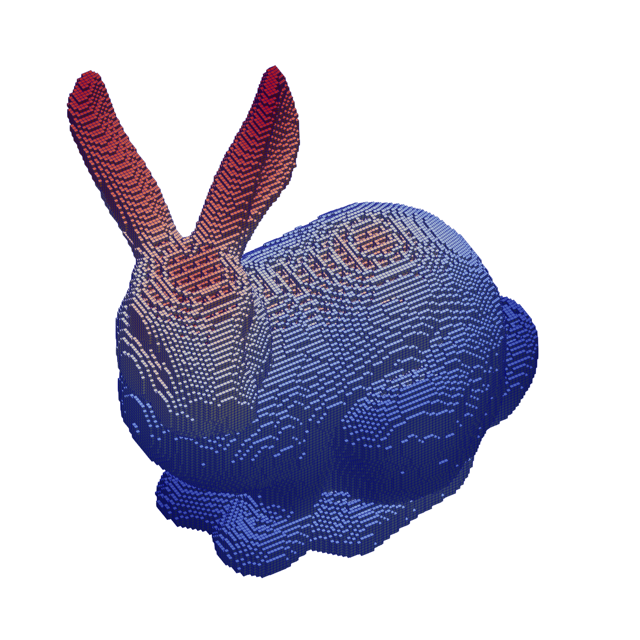}};

        \node[anchor=north west] (vol_0p0_bunny) at (0,1)
            {\includegraphics[width=0.16\linewidth]{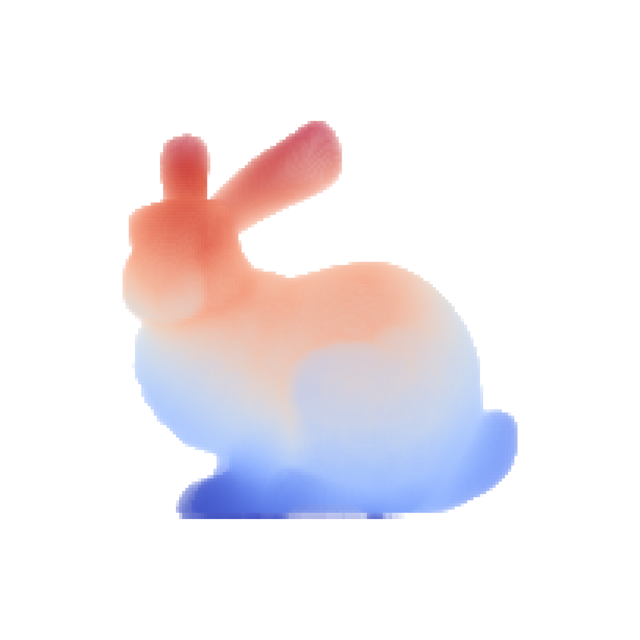}};
        \node[anchor=north west] (vol_0p3_bunny) at (1,1)
            {\includegraphics[width=0.16\linewidth]{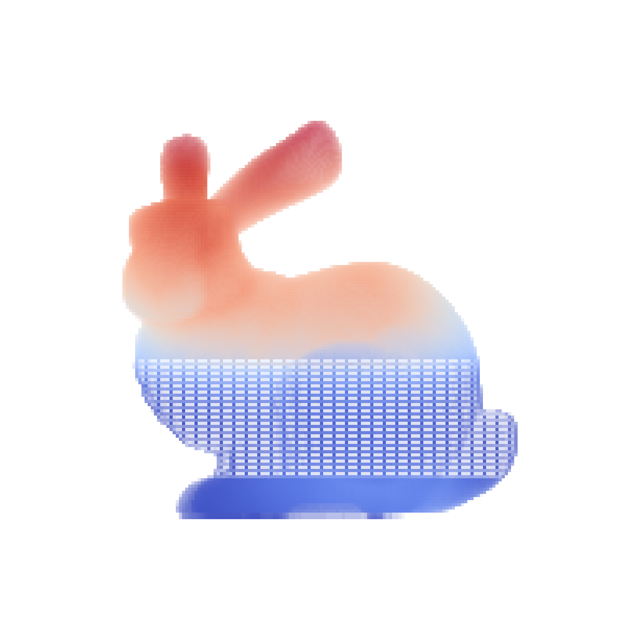}};
        \node[anchor=north west] (vol_0p5_bunny) at (2,1)
            {\includegraphics[width=0.16\linewidth]{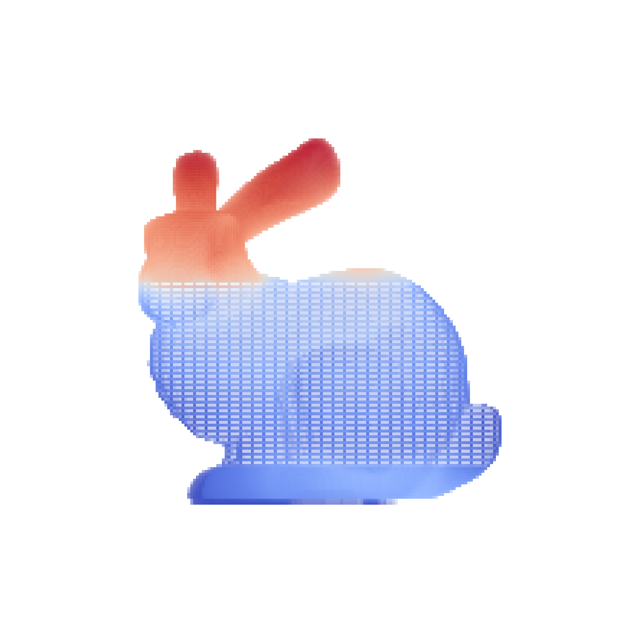}};
        \node[anchor=north west] (vol_0p7_bunny) at (3,1)
            {\includegraphics[width=0.16\linewidth]{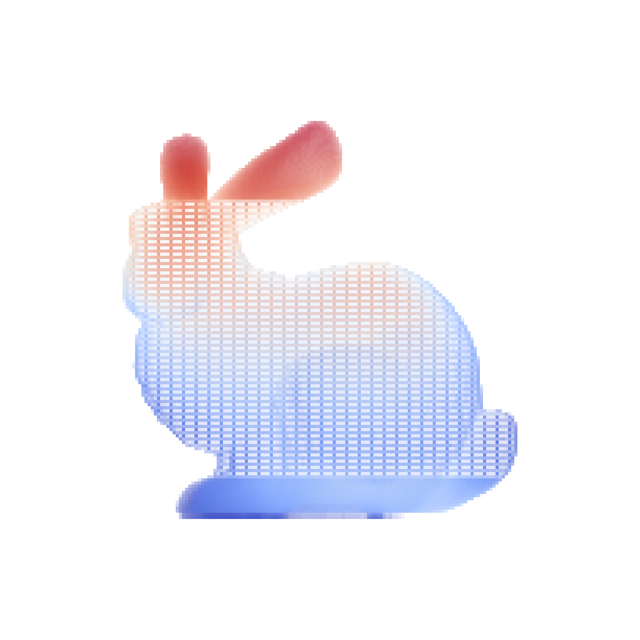}};

        \node[above=0.1cm] at (surf_0p0_bunny.north) {0\%};   
        \node[above=0.1cm] at (surf_0p3_bunny.north) {30\%};    
        \node[above=0.1cm] at (surf_0p5_bunny.north) {50\%};   
        \node[above=0.1cm] at (surf_0p7_bunny.north) {70\%};   

        \node[left=0.1cm] at (surf_0p0_bunny.west)  {Surface View};
        \node[left=0.1cm] at (vol_0p0_bunny.west)  {Volume View};

        \node[below=0.1cm] at (vol_0p3_bunny.south) {\includegraphics[width=6cm, trim={9cm 16cm 8cm 12cm}, clip]{Figures/kelvin_abs_scale.pdf}};

    \end{tikzpicture}
    \caption{Surface and volumetric views of the Stanford Bunny model with different infill sparsities. The top row shows the surface temperature distribution, while the bottom row highlights the volumetric infill pattern. Increased sparsity slows overall heat diffusion and retains higher temperatures in recently deposited layers.}
    \label{Fig:infillSparsity}
\end{figure*}

\figref{Fig:infillSparsity} compares three cases: no sparsity (fully dense), medium sparsity (50\% layers skipped), and high sparsity (75\% layers skipped). As expected, the temperature distribution differs among these configurations, with increased sparsity reducing conduction pathways and causing local thermal gradients to persist longer. This behavior can be relevant for parts where internal voids or lattice structures are desired.

Accurate heat exchange across polymer–air interfaces is achieved with two \emph{distinct} Robin coefficients:
\begin{itemize}
  \item \(h_{\text{infill}}\) on \textbf{internal faces} between an active voxel (polymer) and an inactive voxel (air void),
  \item \(h_{\text{out}}\) on the \textbf{external surface} in contact with the build-chamber air.
\end{itemize}

The coefficients come from classical free-convection theory \citep{incropera2011fundamentals,bejan2013convection} and are summarised in \tabref{Tab:HT_pars_rev}.

\smallskip
\noindent
\textbf{(i) Internal polymer–air gaps (\(h_{\text{infill}}\)).} The narrowest channel width is \(S = 1\;\mathrm{mm}\), while the vertical length scale of a printed layer stack is the body height \(L = 40\;\mathrm{mm}\). Following the vertical–slot analysis of \citet{elenbaas1942}, the Rayleigh number is based on the gap width

\begin{equation}
\text{Ra}_S = \frac{g\,\beta\,(T_s-T_\infty)\,S^{3}}
                   {\nu\,\alpha}=2.44\ (<1708),
\end{equation}

so buoyancy is suppressed and the average Nusselt number takes its pure–conduction limit, \(\text{Nu}=1\). This implies that the internal convection coefficient is simply

\[
h_{\text{infill}}
  =\frac{k\,\text{Nu}}{L}
  =\frac{0.036}{0.04}\;{\rm W\,m^{-2}K^{-1}}
  \approx 0.91\;{\rm W\,m^{-2}K^{-1}} .
\]

\smallskip
\noindent
\textbf{(ii) Outer free-convection film (\(h_{\text{out}}\)).} For outer walls we adopt the Churchill–Chu correlation for a vertical isothermal plate \citep{churchill1975}, valid up to \(\text{Ra}_L\le 10^{12}\):

\begin{align}
\text{Ra}_L &=
      \frac{g\,\beta\,(T_s-T_\infty)\,L^{3}}
           {\nu\,\alpha}=1.56\times 10^{5},\\[2pt]
\mathrm{Nu}_L
  &= \Bigl[\,0.825
           + \frac{0.387\,\mathrm{Ra}_L^{1/6}}
                  {\bigl(1 + (0.492/Pr)^{9/16}\bigr)^{8/27}}
    \Bigr]^{2}
  = 10.24,\quad 10^{4} \le \mathrm{Ra}_L \le 10^{12},\\[2pt]
h_{\text{out}} &= k\,\text{Nu}_L/L
               \approx 9.3\;{\rm W\,m^{-2}K^{-1}},
\end{align}

\begin{table}[t]
  \centering
  \caption{Thermophysical parameters and resulting convection coefficients.}
  \label{Tab:HT_pars_rev}
  \setlength{\extrarowheight}{2pt}
  \begin{tabular}{|l|c|c|}
    \hline
    \textbf{Quantity} & \textbf{Symbol / formula} & \textbf{Value} \\
    \hline\hline
    Geometry height & \(L\) (m) & 0.04 \\
    Gap width & \(S\) (m) & 0.001 \\
    Wall temperature & \(T_s\) (K) & 528.15 \\
    Ambient temperature & \(T_\infty\) (K) & 368.15 \\
    Film temperature & \(T_f\) (K) & 448.15 \\
    Thermal expansion & \(\beta\) (K\(^{-1}\)) & \(2.23\times10^{-3}\) \\
    Dynamic viscosity\textsuperscript{\dag} & \(\mu\) (Pa\,s) & \(2.48\times10^{-5}\) \\
    Density (ideal gas) & \(\rho\) (kg\,m\(^{-3}\)) & 0.79 \\
    Thermal conductivity & \(k\) (W\,m\(^{-1}\)K\(^{-1}\)) & 0.036 \\
    Diffusivity & \(\alpha\) (m\(^{2}\)s\(^{-1}\)) & \(4.57\times10^{-5}\) \\
    Prandtl number & \(Pr\) & 0.69 \\
    \hline
    \multicolumn{3}{|c|}{\emph{Internal gap}} \\ \hline
    Rayleigh number & \(\text{Ra}_S\) & 2.44 \\
    Nusselt number & \(\text{Nu}\) & \(1\) (conduction) \\
    \textbf{Convection coeff.} & \(h_{\text{infill}}=k/L\) & \textbf{0.91} \\
    \hline
    \multicolumn{3}{|c|}{\emph{External wall}} \\ \hline
    Rayleigh number & \(\text{Ra}_L\) & \(1.56\times10^{5}\) \\
    Nusselt number\textsuperscript{\ddag} & \(\text{Nu}_L\) & 10.24 \\
    \textbf{Convection coeff.} & \(h_{\text{out}}\) & \textbf{9.3} \\
    \hline
  \end{tabular}

  \vspace{2pt}
  \footnotesize
  \raggedright
  \textsuperscript{\dag} \citet{kadoya1985}.\;
  \textsuperscript{\ddag}Churchill–Chu correlation \citep{churchill1975}.
\end{table}

The solver therefore applies:
\begin{itemize}
  \item \(h_{\text{infill}}\) on every face shared by an active and an inactive voxel (polymer–void) in the interior,
  \item \(h_{\text{out}}\) on outer faces,
\end{itemize}
ensuring that heat conduction through stagnant air is fully represented while matching experimentally observed external cooling rates.

\begin{figure}[t]
  \centering
  \begin{tikzpicture}
    \begin{axis}[
      width=0.65\linewidth, height=7cm,
      xlabel={Time (s)}, ylabel={Temperature (K)},
      xmin=260, xmax=1200, ymin=340, ymax=560,
      grid=major, grid style={dashed,gray!30},
      legend style={font=\scriptsize,draw=none,
                    at={(0.97,0.97)}, anchor=north east}]
      \addplot[color=orange,    very thick]
        table[x index=5,y index=0,col sep=comma]
        {Data/probe_bunny_0p3.csv};
      \addlegendentry{30\,\% sparse}
      \addplot[color=green!60!black, very thick]
        table[x index=5,y index=0,col sep=comma]
        {Data/probe_bunny_0p5.csv};
      \addlegendentry{50\,\% sparse}
      \addplot[color=red,       very thick]
        table[x index=5,y index=0,col sep=comma]
        {Data/probe_bunny_0p7.csv};
      \addlegendentry{70\,\% sparse}
    \end{axis}
  \end{tikzpicture}
  \caption{Temperature history at an interior probe $(0.0141,\,0.0182,\,0.0069)\,$m for three sparsity levels (i.e. 30\%, 50\%, 70\% sparse layer cases). Higher sparsity earlier the temperature inflection, followed by slow cooling.}
  \label{Fig:infillProbe}
\end{figure}

The probe curves in Fig.~\ref{Fig:infillProbe} display two regimes. For the first \mbox{$\sim\!250$ s} the layers deposited above the probe are \textit{sparse}; they contain little polymer and expose a large internal surface initially to external chamber and later to the voids, so they cool quickly and the probe temperature drops monotonically.  When the toolpath reaches the subsequent \textit{dense} roof layers the situation reverses: the exposed surface area shrinks, convective losses are reduced, and the extra mass stores heat.  Thermal energy from these hotter dense layers diffuses downward, slowing the local cooling rate for all cases or even causing a temporary reheating (30\% case). Although experimental confirmation for such complex cavities is non-trivial, the inflection captured by the model is consistent with the expected thermal behaviour of alternating sparse and roof segments.  A dedicated in-situ measurement campaign is planned as future work.

\subsection{Physical 3D Prints of Complex Geometries}
\label{SubSec:PhysicalPrints}

In order to demonstrate the real-world feasibility and relevance of our thermal simulation framework, we fabricated actual 3D prints of both the Stanford Bunny and the 3D Benchy at sizes matching those used in our simulations. By employing the same voxel-by-voxel deposition strategy, layer heights, and nozzle paths, we aim to show that our simulation predictions are not purely hypothetical but correspond closely to observable outcomes in physical prints. In this section, we present the printer details, geometry dimensions, and material specifications, as well as a direct comparison of printing times and simulation runtimes. Finally, we highlight qualitative observations about the printed parts and discuss how these observations align with our simulation results.

\paragraph{3D Printer and Material Details}
All models were produced on a MakerGear M2 3D printer using ABS filament. Relevant printer specifications and typical operating parameters are summarized in \tabref{Tab:PrinterSpecs}. The bed and nozzle temperatures were set to $373.15K$ and $513.15K$, respectively, to optimize adhesion and minimize warping. The nozzle diameter was set to $0.35\,mm$, and the layer height was set to match the voxel height used in our simulations ($0.28\,mm$).

\begin{table}[ht!]
    \centering
    \caption{Basic 3D printer parameters (MakerGear M2) used for fabricating the Bunny and Benchy models.}
    \label{Tab:PrinterSpecs}
    \setlength{\extrarowheight}{3pt}
    \begin{tabular}{|l|l|}
    \hline
    \textbf{Printer Model} & MakerGear M2 \\
    \textbf{Filament} & ABS \\
    \textbf{Nozzle Diameter} & $0.35\,mm$ \\
    \textbf{Layer Height} & $0.28\,mm$ \\
    \textbf{Bed Temperature} & $373.15K$ \\
    \textbf{Nozzle Temperature} & $513.15K$ \\
    \hline
    \end{tabular}
\end{table}

\paragraph{Geometry Details}
\tabref{Tab:GeomPrints} outlines the dimensions and voxelization information for the Stanford Bunny and 3D Benchy. Notably, the layer height on the physical printer is set equal to the voxel height used in our simulations, ensuring a consistent basis for comparing geometry size, build times, and layer-by-layer formation. These geometrie's bounding boxes match those in our simulations, and the nominal voxel sizes (along all three axes) directly correspond to the physical layer height and in-plane path spacing used for printing. Because no hardware constraints prevented matching the simulation layer height, there was no discrepancy between the printed layer thickness and the voxel height in this work.

\begin{table}[t!]
    \centering
    \setlength{\extrarowheight}{3pt}
    \caption{Geometry details for the Stanford Bunny and 3D Benchy prints. ``Voxels'' and ``Voxel Size'' match the resolution used in the thermal simulations.}
    \label{Tab:GeomPrints}
    \begin{tabular}{|c|c|c|c|}
    \hline
    \textbf{Model} & \textbf{Bounding Box} & \textbf{Voxels} & \textbf{Voxel Size} \\
    \hline
    Bunny &  $ 28.2 \times 36.2 \times 35.9 $ & $464158$ & $ 0.28 \times 0.28 \times 0.28 $\\
    Benchy & $19.1 \times 36.5 \times 29.3$ & $175481$ & $ 0.28 \times 0.29 \times 0.28 $\\
    \hline
    \end{tabular}
\end{table}

\paragraph{Comparison of Simulation Time with Print Time}
A key goal of our work is to show that the proposed simulation can run \emph{faster} than the actual 3D print time, even at the same resolution and approximate deposition rate. \tabref{Tab:TimeComparison} compares the total print times versus the total simulation times for both the Bunny and Benchy prints. Notably, the Bunny simulation requires around \textbf{48 minutes} to complete, while the actual print finishes in roughly \textbf{148 minutes}. Similarly, the Benchy simulation finishes in about \textbf{55 minutes} compared to the actual print time of \textbf{96 minutes}. These results underscore our framework’s faster-than-real-time predictive capability, thereby opening avenues for in-situ monitoring or pre-print optimization.

\begin{table}[t!]
    \centering
    \renewcommand{\arraystretch}{1.15}
    \caption{Comparison between simulation runtimes and actual 3D-printing durations for each geometry at matching scales.}
    \label{Tab:TimeComparison}
    \begin{tabular}{|l|c|c|}
    \hline
    & \textbf{Simulation Time (mins)} & \textbf{Print Time (mins)} \\
    \hline
    \textbf{Bunny}  & $48$ & $148$ \\
    \textbf{Benchy} & $55$ & $96$ \\
    \hline
    \end{tabular}
\end{table}

\begin{figure}[!t]
    \centering
    \includegraphics[width=0.38\linewidth,clip,trim={0in 1in 0in 2in}]{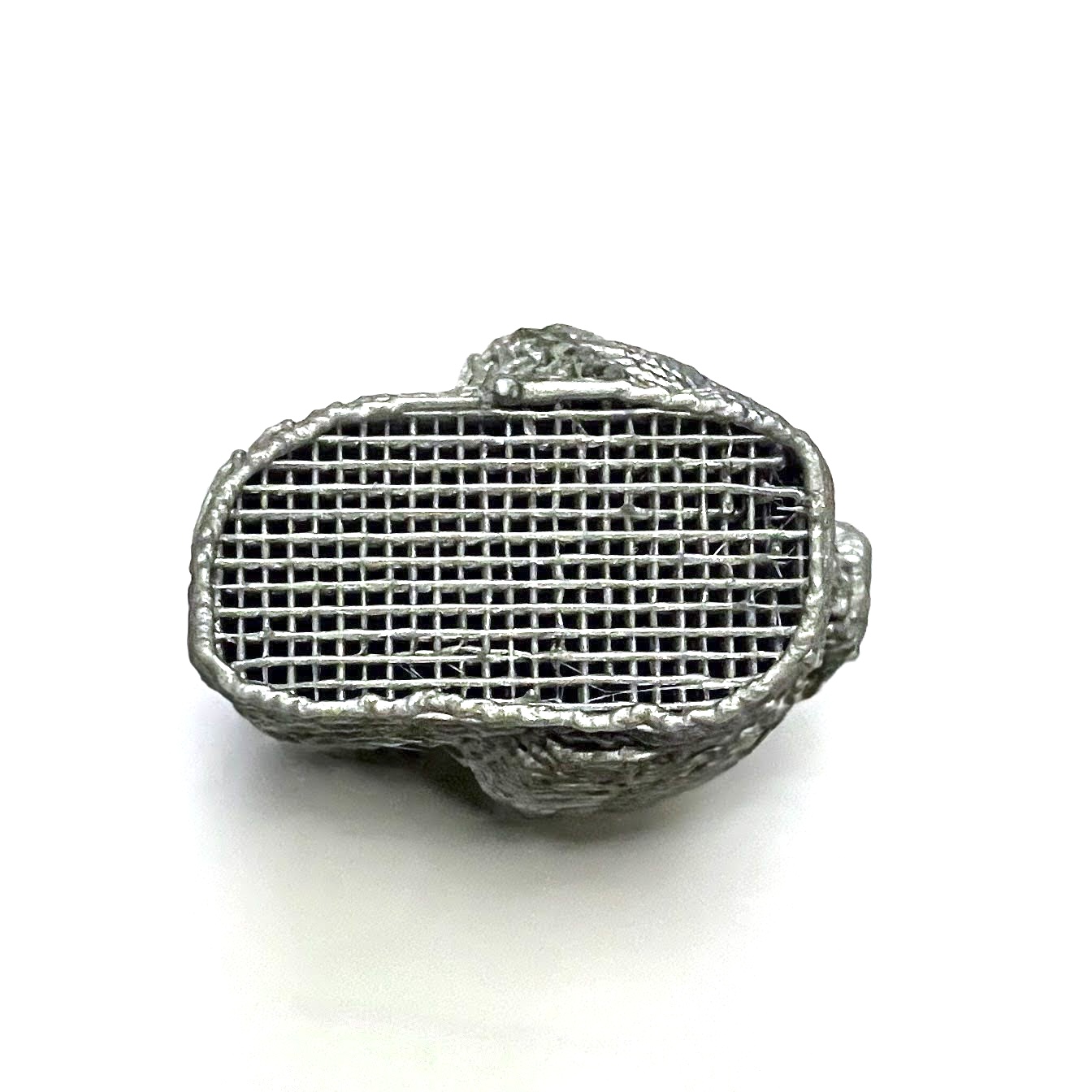}
    \caption{Top view of the bunny model printed up to an intermediate height showing the voxel pattern.}
    \label{Fig:topViewIntermediate}
\end{figure}

\begin{figure*}[!t]
    \centering
    \begin{subfigure}{0.24\linewidth}
        \centering
        \includegraphics[width=\linewidth]{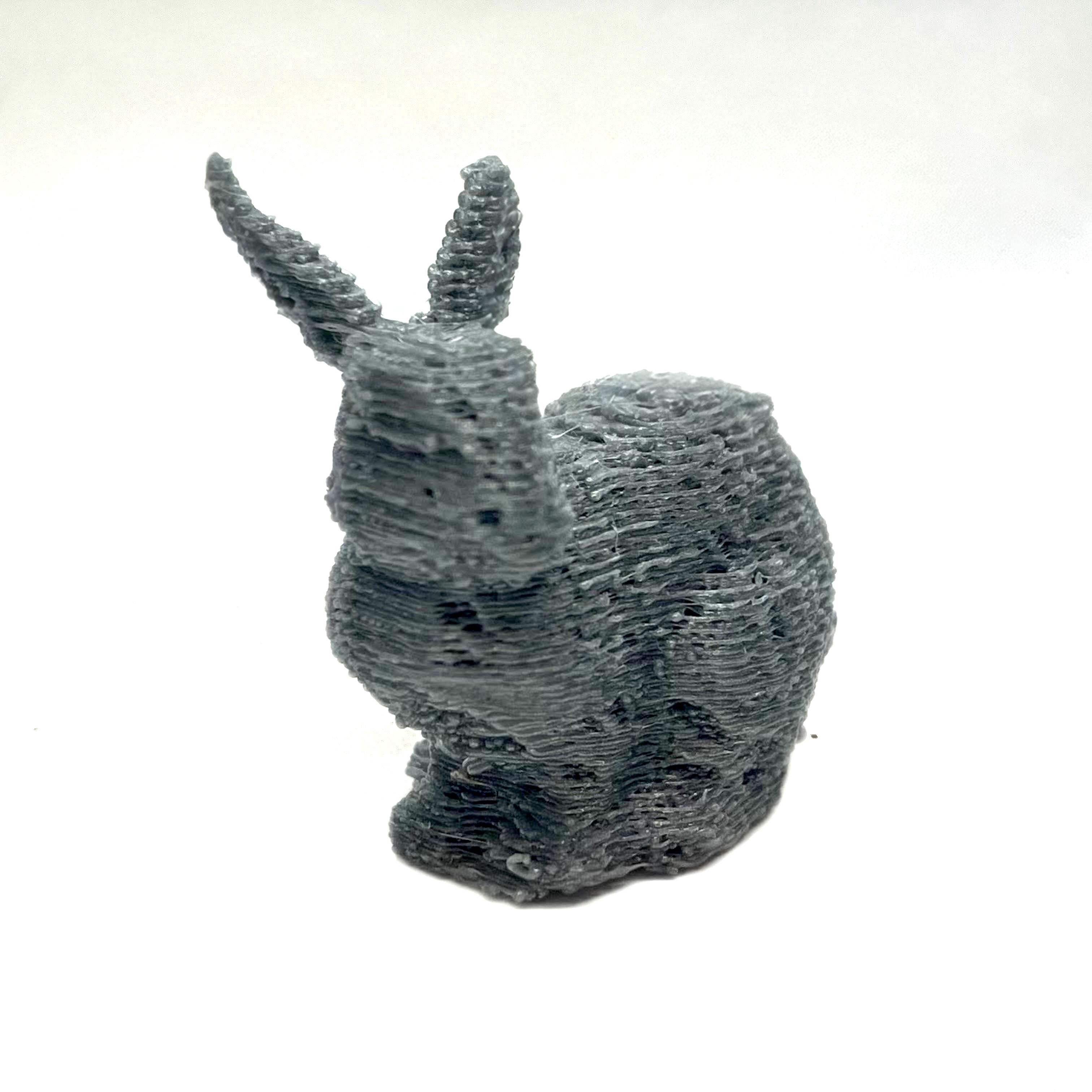}
        \caption{Stanford Bunny view 1 (ABS)}
    \end{subfigure}
    \hfill
    \begin{subfigure}{0.24\linewidth}
        \centering
        \includegraphics[width=\linewidth]{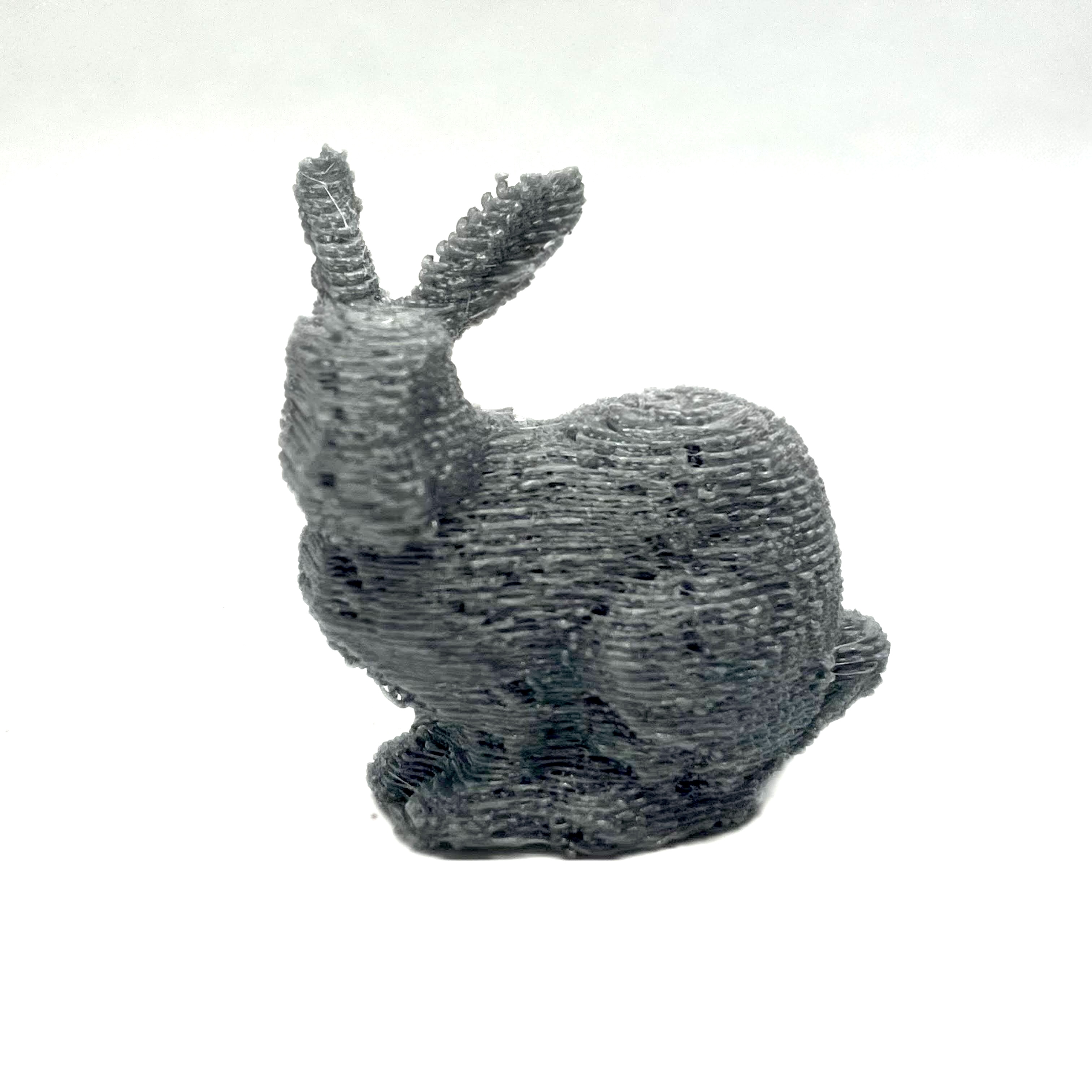}
        \caption{Stanford Bunny view 2 (ABS)}
    \end{subfigure}
    \hfill
    \begin{subfigure}{0.24\linewidth}
        \centering
        \includegraphics[width=\linewidth]{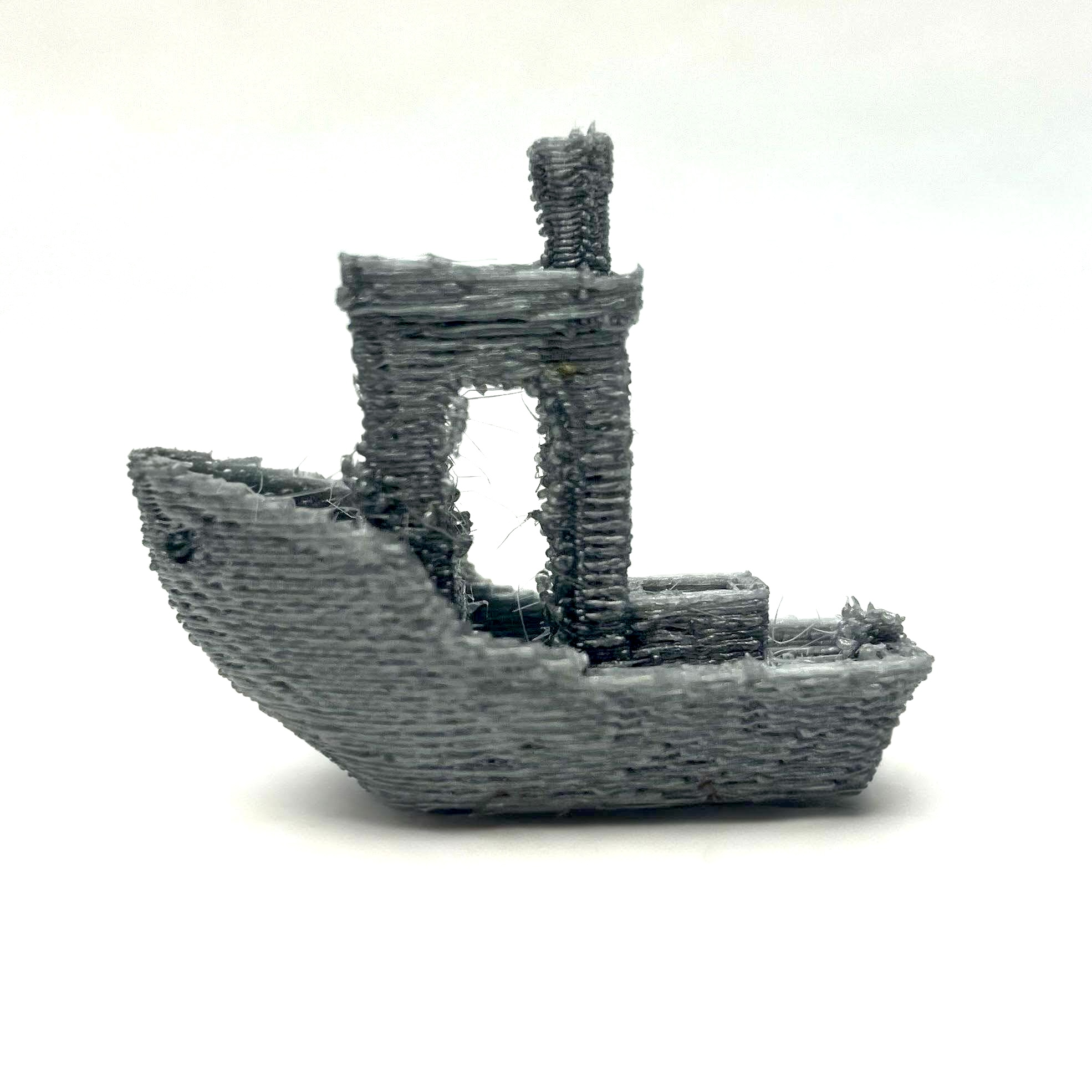}
        \caption{3D Benchy view 1 (ABS)}
    \end{subfigure}
    \hfill
    \begin{subfigure}{0.24\linewidth}
        \centering
        \includegraphics[width=\linewidth]{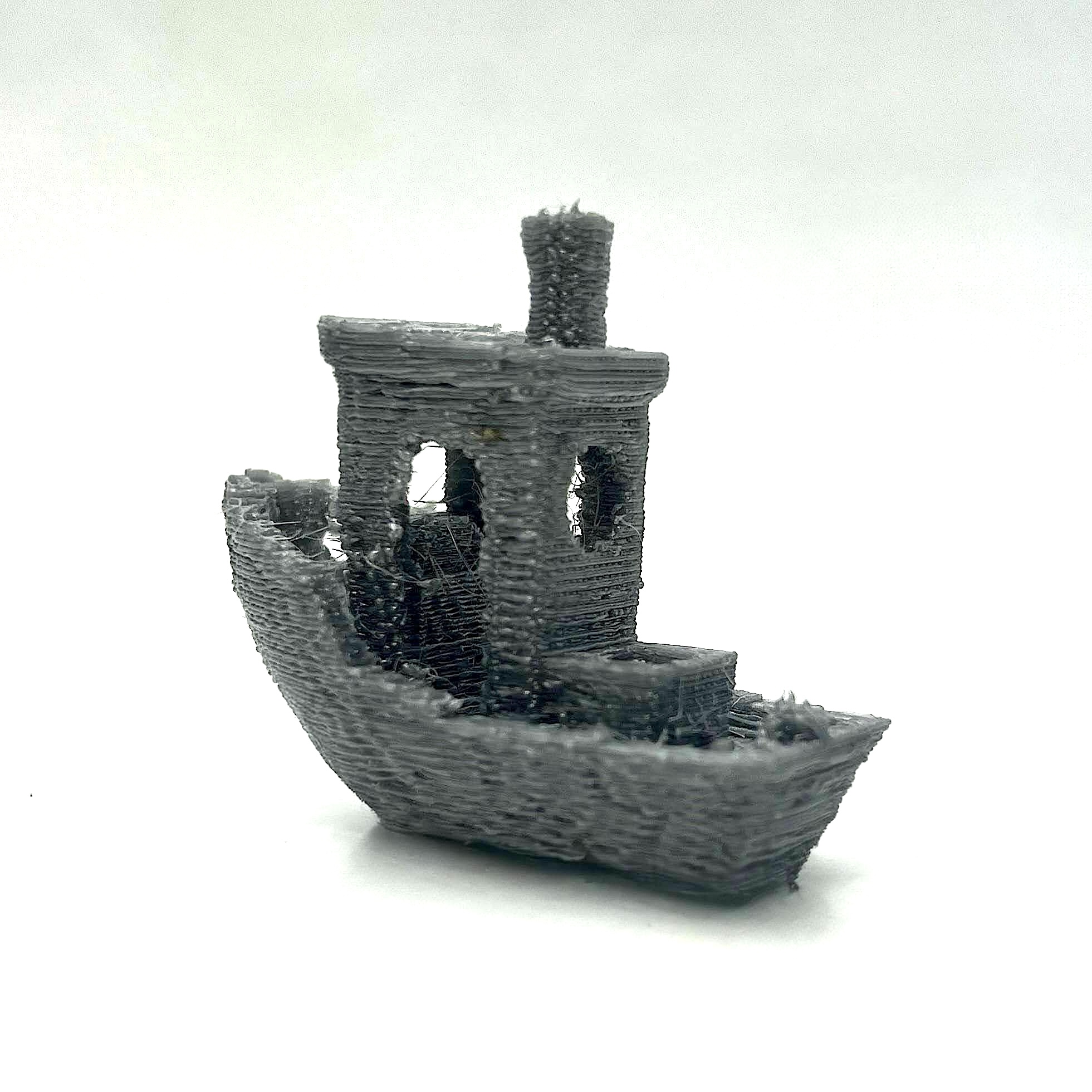}
        \caption{3D Benchy view 2 (ABS)}
    \end{subfigure}
    \caption{Photographs of actual 3D prints corresponding to our simulations. Fine surface details and minimal warping indicate that the thermal gradients are well-managed for these geometries.}
    \label{Fig:3DprintedModels}
\end{figure*}

\paragraph{Qualitative correspondence between prints and simulation}
The photographs in \figref{Fig:topViewIntermediate}–\ref{Fig:3DprintedModels}
serve a \emph{qualitative} purpose: they demonstrate that a voxel‐by‐voxel
(G-code–driven) tool path that is \emph{identical} to the one used in the
thermal solver can be fabricated successfully on a commodity printer, even for
complex, sparse-infill shapes.  The absence of severe warping, layer
delamination or large surface defects is \emph{consistent} with the moderate
thermal gradients predicted by our model for the chosen print parameters.
We emphasise, however, that these images are \emph{not} put forward as a
quantitative validation of the heat-transfer solution inside the infill
region.  A rigorous validation would require time-resolved infrared
thermography or embedded thermocouples—an experimental campaign that lies
outside the scope of the present, framework-oriented study.  Instead, the
prints demonstrate (i) that the voxel geometry exported by our pipeline is
faithfully realised in hardware and (ii) that the simulation can be executed
faster than the real build, opening the door to future in-situ correction
schemes.  Quantitative temperature measurements in sparsely filled cavities
are planned as part of our follow-up work.

\section{Discussion and Conclusions}
\label{Sec:DiscussionConclusion}

The results presented in this paper underscore both the accuracy and computational efficiency of our finite element framework for thermal simulations of extrusion-based additive manufacturing. We validated our method with single-filament experiments for ABS and PEKK (\secref{SubSec:SingleFilamentWall}), demonstrating that our voxel-based approach accurately captures the transient temperature evolution with minimal runtime. The comparisons to reference experimental and numerical data \citep{lepoivre2020heat,nagaraj2023novel} confirm that fundamental heat transfer phenomena---particularly conduction within and between deposited filaments, as well as convective losses to the surrounding environment---are effectively modeled at scales directly relevant to typical 3D-printing processes.

Building on these foundational results, we have extended the framework to handle more complex geometries---the Stanford Bunny, 3D Benchy, and Moai (\secref{SubSec:ComplexGeometrySimulations}). Our simulations demonstrated stable performance across multiple voxel resolutions (up to $128^3$), showcasing the method’s capability to capture intricate layer-by-layer thermal behavior for geometries of practical interest in additive manufacturing. Beyond purely geometric complexity, we also investigated the influence of sparse infill patterns on temperature distribution (\secref{SubSec:SparselyInfillGeometries}), finding that elevated local temperatures persist in regions with fewer conduction pathways. This insight can guide the design of lightweight or lattice-filled components by identifying where potential thermal-driven defects---such as local overheating or interlayer adhesion issues---are more likely to occur.

An important benefit of our approach is that it is faster than actual print time performance, as demonstrated by the simulation runtimes for the Bunny and Benchy models (\secref{SubSec:PhysicalPrints}). This capability opens up exciting possibilities for real-time process monitoring or pre-print optimization, where manufacturers can adjust parameters on-the-fly based on predicted thermal behavior. Our framework can adaptively refine or coarsen an octree-based mesh at runtime, which opens the door to high-resolution simulations over large build volumes without prohibitive computational costs. This flexibility can be crucial for industrial applications, where both geometric fidelity and throughput are paramount.

There are several research opportunities that can be explored in the future. First, the inclusion of fully coupled thermo-mechanical models would enhance predictions of warping, residual stress, and layer-to-layer bonding---phenomena critical to part reliability. Second, more complex boundary conditions (e.g., forced convection in enclosed chambers) could further improve the fidelity of predictions for industrial-grade printers and exotic materials. Third, exploring the temperature dependence of material properties, along with crystallization kinetics in high-performance polymers, may refine thermal predictions in extreme build environments. Addressing these challenges would expand the framework’s applicability and further embed it as a robust, versatile tool for process simulation.

In conclusion, our voxel-based FEM approach offers an efficient, scalable, and accurate solution for modeling heat transfer in extrusion-based additive manufacturing. The framework has been rigorously validated against established experimental and numerical benchmarks and can handle intricate geometries, varying infill strategies, and a wide range of material systems. This framework demonstrates the potential to provide predictive capabilities on timescales that are either faster than or comparable to the actual printing time. Such capabilities are significant for real-time process monitoring, adaptive control, and virtual prototyping applications---ultimately helping manufacturers optimize print quality, reduce trial-and-error cycles, and bring advanced additive manufacturing applications closer to widespread adoption.

\section*{Acknowledgements}
This work was supported in part by the National Science Foundation under grant numbers LEAP-HI-2053760, CMMI-2347623, DMREF-2323716 and by the Office of the Under Secretary of Defense for Research and Engineering, Strategic Technology Protection and Exploitation, and Defense Manufacturing Science and Technology Program under agreement number W15QKN-19-3-0003. The U.S. Government is authorized to reproduce and distribute reprints for governmental purposes, notwithstanding any copyright notation thereon. The views and conclusions expressed in this work are those of the authors and do not necessarily reflect the official policies or endorsements of the U.S. Government.

\bibliographystyle{elsarticle-num-names}
\bibliography{Refs}

\newpage
\appendix

\begin{center}
{
{\usefont{OT1}{phv}{b}{n}\selectfont\Large{Supplementary Information}}
\vspace{0.5em}
}
\end{center}

\section{Hardware Setup and Computing Environments}
\label{Sec:HardwareSetup}

We performed simulations on two primary systems:

\begin{itemize}
    \item \textbf{Workstation for Single-Filament Study.} For the single-filament wall simulations (ABS and PEKK) \secref{SubSec:SingleFilamentWall}, we used a standard workstation equipped with an \textbf{Intel(R) Core(TM) i7-4790 CPU @ 3.60GHz}. \tabref{Tab:single_filament_cpu_specs} summarizes the key hardware details.

    \item \textbf{Workstation for Complex Geometry and Sparse Infill.} 
    For the larger-scale Bunny, Benchy, and Moai simulations (\secref{SubSec:ComplexGeometrySimulations}, 
    \secref{SubSec:SparselyInfillGeometries}) we utilized more powerful 56~CPUs based HPC cluster environment, we leveraged one or more nodes of an \textbf{Intel(R) Xeon(R) Platinum 8280 CPU @ 2.70GHz} architecture. Key CPU specifications are listed in \tabref{Tab:complex_geo_cpu_specs}. 
\end{itemize}

\begin{table}[!ht]
    \centering
    \renewcommand{\arraystretch}{1.15}
    \begin{tabular}{|l|l|}
    \hline
    \textbf{Hardware Parameter} & \textbf{Value}\\
    \hline
    CPU model & Intel(R) Core(TM) i7-4790 @ 3.60GHz \\
    \hline
    Cores/Threads & 4 cores, 8 threads total \\
    \hline
    L3 cache & 8\,MB \\
    \hline
    Max CPU frequency & 4.0\,GHz \\
    \hline
    \end{tabular}
    \caption{Hardware details for the single-filament wall simulations in \secref{SubSec:SingleFilamentWall}.}
    \label{Tab:single_filament_cpu_specs}
\end{table}

\begin{table}[!ht]
    \centering
    \renewcommand{\arraystretch}{1.15}
    \begin{tabular}{|l|l|}
    \hline
    \textbf{Hardware Parameter} & \textbf{Value}\\
    \hline
    CPU model & Intel(R) Xeon(R) Platinum 8280 @ 2.70GHz \\
    \hline
    Cores/Threads & 28 cores per socket, 2 sockets, 56 CPUs total \\
    \hline
    L3 cache & 39424\,KB \\
    \hline
    Max CPU frequency & 4.0\,GHz \\
    \hline
    \end{tabular}
    \caption{Hardware details for the complex geometry and sparse-infill simulations in \secref{SubSec:ComplexGeometrySimulations} and \secref{SubSec:SparselyInfillGeometries}.}
    \label{Tab:complex_geo_cpu_specs}
\end{table}

We report the relevant CPU, cache, and memory layout for completeness, ensuring that readers can reproduce or benchmark similar setups if desired. 
Details on the specific number of MPI processes or threads utilized in each simulation are given in \secref{SubSubSec:ComplexGeoSimParams} and \secref{SubSec:SparselyInfillGeometries}.

\section{Supplementary Video}
\label{SubSec:SupplementaryVideo}

The full thermal simulation video for the $128^3$ voxel resolution Stanford bunny, Benchy and Moai during the 3D printing process is attached as supplementary material. The video shows the $VbV$ 3D printing process along with the temperature distribution, the variation in the infill sparsity with model height can also be seen. These visualizations show that the physics simulations can capture the complex interaction between the 3D printing parameters and the final printed object.

\end{document}